\documentclass[journal,twocolumn,10pt]{IEEEtran}
\usepackage{graphicx}
\usepackage{amssymb}
\usepackage{amsmath}
\usepackage{mathtools}
\usepackage{dsfont}
\usepackage{cite}
\usepackage{stfloats}
\usepackage{subfigure}
\usepackage{psfrag}
\usepackage[mathscr]{euscript}
\usepackage{acronym}  
\usepackage{booktabs}

\usepackage{float}
\usepackage{balance}

\acrodef{CCDF}{complementary cumulative distribution function}
\acrodef{CF}{characteristic function}
\acrodef{PPP}{Poisson point processe}
\acrodef{RV}{random variable}
\acrodef{i.i.d.}{independent and identically distributed}
\acrodef{PDF}{probability distribution function}
\acrodef{CDF}{cumulative distribution function}
\acrodef{ch.f.}{characteristic function}
\acrodef{AWGN}{additive white Gaussian noise}
\acrodef{SNR}{signal-to-noise ratio}
\acrodef{LRT}{likelihood ratio test}
\acrodef{DRT}{distance ratio test}
\acrodef{GLRT}{generalized likelihood ratio test}
\acrodef{CRLB}{Cram\'{e}r-Rao lower bound}
\acrodef{CRB}{Cram\'{e}r-Rao bound}
\acrodef{ZZLB}{Ziv-Zakai lower bound}
\acrodef{ZZB}{Ziv-Zakai bound}
\acrodef{LOS}{line-of-sight}
\acrodef{ToF}{time-of-flight}
\acrodef{NLOS}{non-line-of-sight}
\acrodef{GDOP}{geometric dilution of precision}
\acrodef{GPS}{Global Positioning System}
\acrodef{FIM}{Fisher information matrix}
\acrodef{PEB}{position error bound}
\acrodef{SPEB}{squared position error bound}
\acrodef{TOA}{time-of-arrival}
\acrodef{TOF}{time-of-flight}
\acrodef{WSN}{wireless sensor network}
\acrodef{MAC}{medium access control}
\acrodef{RSS}{received signal strength}
\acrodef{WAF}{wall attenuation factor}
\acrodef{TDOA}{time difference-of-arrival}
\acrodef{RF}{radiofrequency}
\acrodef{RTT}{round-trip time}
\acrodef{AOA}{angle-of-arrival}
\acrodef{MF}{matched filter}
\acrodef{ED}{energy detector}
\acrodef{ML}{maximum likelihood}
\acrodef{MSE}{mean-square error}
\acrodef{RMSE}{root-mean-square error}
\acrodef{LEO}{localization error outage}
\acrodef{ppm}{part-per-million}
\acrodef{ACK}{acknowledge}
\acrodef{UWB}{Ultrawide bandwidth}
\acrodef{TNR}{threshold-to-noise ratio}
\acrodef{LS}{least squares}
\acrodef{IR-UWB}{impulse radio UWB}
\acrodef{FCC}{Federal Communications Commission}
\acrodef{TH}{time-hopping}
\acrodef{PPM}{pulse position modulation}
\acrodef{MUI}{multi-user interference}
\acrodef{PDP}{power delay profile}
\acrodef{BPZF}{band-pass zonal filter}
\acrodef{SIR}{signal-to-interference ratio}
\acrodef{SINR}{signal-to-interference-plus-noise ratio}
\acrodef{RFID}{radio frequency identification}
\acrodef{WPAN}{wireless personal area network}
\acrodef{WWB}{Weiss-Weinstein bound}
\acrodef{DP}{direct path}
\acrodef{MF}{matched filter}
\acrodef{MMSE}{minimum-mean-square-error}
\acrodef{SBS}{serial backward search}
\acrodef{SBSMC}{serial backward search for multiple clusters}
\acrodef{NBI}{narrowband interference}
\acrodef{WBI}{wideband interference}
\acrodef{INR}{interference-to-noise ratio}
\acrodef{CR}{channel response}
\acrodef{CIR}{channel impulse response}
\acrodef{CR}{channel  response}
\acrodef{RADAR}{radar}
\acrodef{MUR}{Multistatic radar}
\acrodef{JBSF}{jump back and search forward}
\acrodef{HDSA}{high-definition situation-aware}
\acrodef{RRC}{root raised cosine}
\acrodef{ST}{simple thresholding}
\acrodef{BTB}{Bellini-Tartara bound}
\acrodef{P-Max}{$P$-Max}  
\acrodef{MIMO}{multiple-input multiple-output}
\acrodef{MAP}{maximum a posteriori}
\acrodef{FG}{factor graph}
\acrodef{OP}{outage probability}
\acrodef{WED}{wall extra delay}
\acrodef{RMS}{root mean square}
\acrodef{SPAWN}{sum-product algorithm over a wireless network}
\acrodef{MDD}{minimum distance distribution}
\acrodef{MAP}{maximum a posteriori probability}
\acrodef{SAP}{small cell access point}
\acrodef{UE}{user equipment}
\acrodef{MBS}{macro cell base station}
\acrodef{UER}{\ac{UE} Relay}
\acrodef{D2D}{device-to-device}
\acrodef{MBS}{macro base station}
\acrodef{CSI}{channel state information}
\acrodef{OGR}{outage guard region}
\acrodef{FUR}{feasible UER region}
\acrodef{EHR}{energy harvesting region}
\acrodef{EH}{energy harvesting}
\acrodef{D2D-EHSN}{D2D communication provided \ac{EH} small cell network}
\acrodef{D2D-EHHN}{D2D communication provided \ac{EH} heterogeneous network}
\acrodef{3GPP}{3rd Generation Partnership Project}
\acrodef{BS}{base station}
\acrodef{DF}{decode and forward}
\acrodef{CCDF}{complementary cumulative distribution function}
\acrodef{ZF}{zero forcing}
\acrodef{RZF}{regularized zero forcing}
\acrodef{WLLN}{weak law of large number}
\acrodef{SLLN}{strong law of large numbers}
\acrodef{TDD}{Time-division duplex}
\acrodef{EE}{energy efficiency} 
\acrodef{HetNet}{heterogeneous network} 
\acrodef{SCP}{Single Cell Processing}
\acrodef{CBF}{Coordinated Beamforming}
\usepackage{color}
\usepackage{dsfont}
\usepackage{bbm}

\def\PMT{P_{\mathrm{t}}}

\def\KM{K}

\DeclareMathAlphabet{\mathsf}{OML}{cmbr}{m}{it}

\newtheorem{theorem}{\bf Theorem}
\newtheorem{lemma}{\bf Lemma}
\newtheorem{corollary}{\bf Corollary}
\newtheorem{proposition}{\bf Proposition}

\newtheorem{remark}{\bf Remark}

\newtheorem{assumption}{\bf Assumption}

\newcommand{\bd}{\begin{description}}
\newcommand{\ed}{\end{description}}
\newcommand{\be}{\begin{enumerate}}
\newcommand{\ee}{\end{enumerate}}
\newcommand{\bi}{\begin{itemize}}
\newcommand{\ei}{\end{itemize}}
\newcommand{\bl}{\begin{list}}
\newcommand{\el}{\end{list}}
\newcommand{\bt}{\begin{tabbing}}
\newcommand{\et}{\end{tabbing}}

\newcommand{\paperTitle}{Cell-Edge-Aware Precoding for Downlink\\Massive MIMO Cellular Networks}

\begin{document}

{
\title{\paperTitle}

\author{

	    Howard~H.~Yang, \textit{Student Member, IEEE},
        Giovanni~Geraci, \textit{Member, IEEE}, \\
        Tony~Q.~S.~Quek, \textit{Senior Member, IEEE}, and
        Jeffrey~G.~Andrews, \textit{Fellow, IEEE}
\thanks{H.~H.~Yang and T.~Q.~S.~Quek are with the Singapore University of Technology and Design, Singapore (e-mail: hao\_yang@mymail.sutd.edu.sg, tonyquek@sutd.edu.sg).}
\thanks{G.~Geraci is with Bell Laboratories Nokia, Dublin, Republic of Ireland (e-mail: giovanni.geraci@nokia.com).}
\thanks{J.~G.~Andrews is with the University of Texas at Austin, Austin TX, USA (email: jandrews@ece.utexas.edu).}
    }
\maketitle
\acresetall
\thispagestyle{empty}
\begin{abstract}
We propose a \textit{cell-edge-aware} (CEA) zero forcing (ZF) precoder that exploits the excess spatial degrees of freedom provided by a large number of base station (BS) antennas to suppress inter-cell interference at the most vulnerable user equipments (UEs). We evaluate the downlink performance of CEA-ZF, as well as that of a conventional \textit{cell-edge-unaware} (CEU) ZF precoder in a network with random base station topology. Our analysis and simulations show that the proposed CEA-ZF precoder outperforms CEU-ZF precoding in terms of (i) aggregate per-cell data rate, (ii) coverage probability, and (iii) $95\%$-likely, or edge user, rate.
In particular, when both perfect channel state information and a large number of antennas $N$ are available at the BSs, we demonstrate that the outage probability under CEA-ZF and CEU-ZF decay as ${1}/{N^2}$ and ${1}/{N}$, respectively. This result identifies CEA-ZF as a more effective precoding scheme for massive MIMO cellular networks.
Our framework also reveals  the importance of scheduling the optimal number of UEs per BS, and  confirms the necessity to control the amount of pilot contamination received during the channel estimation phase.
\end{abstract}

\acresetall

\section{Introduction}\label{sec:intro}

Supporting the ever increasing wireless throughput demand is the primary factor driving the industry and academia alike towards the fifth generation (5G)  wireless systems. Not only will 5G cellular networks have to provide a large aggregate capacity, but {more importantly}, they will have to guarantee high worst-case rates for all \acp{UE}, including those located at the cell edge, i.e., close to interfering \acp{BS} \cite{Ericsson:16,Nokia:15,AndBuzCho:14}. New technologies are being introduced to improve the performance of cell-edge \acp{UE} from current levels. Equipping \acp{BS} with a large number of antennas, widely known as massive multiple-input multiple-output (MIMO), has emerged as one of the most promising solutions \cite{Mar:10,RusPerBuoLar:2013,LuLiSwi:14}. In this work, we propose to use some of the spatial dimensions available at massive MIMO \acp{BS} to significantly improve the data rate of \acp{UE} at the cell edge, as well as the overall network throughput. To this end, we design and analyze a linear transmission scheme, termed \textit{cell-edge-aware} (CEA) zero forcing (ZF) precoder, that suppresses interference at the cell-edge \acp{UE}.

\subsection{Motivation and Related Work}

A considerable amount of research has investigated the use of multi-cell joint signal processing for cell-edge performance improvement \cite{NigMinHae:14,XuYanLi:14,RalSinAnd:14}. The common idea behind joint processing techniques is to organize \acp{BS} in clusters, where \acp{BS} lying in the same cluster share information on the data to be transmitted to all \acp{UE} in the cluster. Although this information allows \acp{BS} to coordinate their transmissions and jointly serve all \acp{UE} with an improved system throughput, it comes at the cost of heavy signaling overhead and backhaul latency, which defy the purpose of its implementation \cite{LozHeaAnd:13}.

As the benefits of joint processing are often outweighed by the increased latency and overhead, a more practical alternative to increase the cell-edge throughput can be found in coordinated beamforming, or \textit{precoding}, schemes \cite{ZakHan:12,BhaHea:11,HuaDurZho:15}. Under coordinated precoding, each \ac{BS} acquires additional channel state information (CSI) of \acp{UE} in neighboring cells, but no data information is shared between the various \acp{BS}. The additional CSI can then be exploited to control the crosstalk generated at \acp{UE} in other cells, e.g., by using multiple \ac{BS} antennas to steer the crosstalk towards the nullspace of the neighboring \acp{UE}. This approach is especially attractive for massive MIMO \acp{BS}, due to the abundance of spatial dimensions provided by the large antenna arrays \cite{HoyHosBri:13}.

Recent attempts to design and analyze a coordinated precoder for massive MIMO cellular networks are made in \cite{BjoLarDeb:16,ZhuWanQia:16}. The current paper differs from and generalizes these two works in two key aspects:
\begin{enumerate}
\item \textit{Design}: Unlike \cite{BjoLarDeb:16,ZhuWanQia:16}, where each \ac{BS} suppresses the interference at all edge \acp{UE} in all neighboring cells, we specifically target those neighboring \acp{UE} close to the \ac{BS} coverage area. Therefore, our precoder employs fewer spatial dimensions to mitigate inter-cell interference, leaving more degrees of freedom to each \ac{BS} to better multiplex its own associated \acp{UE} \cite{hoy:13massive}.
\item \textit{Analysis}: While \cite{BjoLarDeb:16,ZhuWanQia:16} assume a symmetric hexagonal cellular network, we consider a {network} model with random topology. Hexagonal models can lead to substantial performance overestimation, as demonstrated in \cite{Fujitsu:11,BjoLar:15}, whereas our analysis accounts for the randomness of practical cellular deployments.
\end{enumerate}

\subsection{Approach and Summary of Results}
In this paper, we model the massive MIMO BS deployment and the UE locations as independent \acp{PPP}, where each BS simultaneously serves multiple UEs on each time-frequency resource block (RB). By introducing a second-order Voronoi tessellation, we define the cell neighborhood of each BS, and design a CEA-ZF precoder that controls the interference generated in the neighborhood. Using random matrix theory and stochastic geometry, we analyze the coverage and rate performance of CEA-ZF, as well as that of a conventional \emph{cell-edge-unaware} (CEU) ZF precoder, in a general setting that accounts for the interference affecting both the channel estimation and data transmission phases.
Our contributions can be summarized as follows.
\begin{itemize}
\item We propose a new precoder for the downlink of massive MIMO cellular networks, which we denote as the CEA-ZF precoder, where some spatial dimensions are used to suppress inter-cell interference at the cell-edge neighboring UEs, and the remaining degrees of freedom are used to multiplex \acp{UE} within the cell. Our precoder works in a distributed manner, and it boosts network coverage and rate performance compared to CEU-ZF precoding.
\item We develop a general framework to analyze the \ac{SIR} distribution and coverage of massive MIMO cellular networks for both the proposed CEA-ZF and the CEU-ZF precoder. Our analysis is tractable and captures the effects of multi-antenna transmission, spatial multiplexing, path loss and small-scale fading, network load and \ac{BS} deployment density, imperfect channel estimation, and random network topology.
\item Through our analysis, which is validated via simulation results, we show that the proposed CEA-ZF precoder outperforms conventional CEU-ZF in terms of aggregate per-cell data rate and coverage probability. Moreover, CEA-ZF guarantees a significantly larger $95\%$-likely rate, i.e., it improves the cell-edge rate. When perfect CSI and a large number of antennas, $N$, are available at the BSs, we find that the outage probabilities under CEA-ZF and CEU-ZF decay as $1/N^2$ and $1/N$, respectively, demonstrating that CEA-ZF is a more effective precoding scheme for massive MIMO cellular networks.
\item We quantify the effect of imperfect CSI, and reveal the importance of controlling the amount of pilot contamination received during the channel estimation phase, e.g., through smart pilot allocation schemes. We also study the system performance as a function of the network load, showing that the aggregate per-cell rate is sensitive to the number of UEs spatially multiplexed on the same RB.
\end{itemize}

The remainder of the paper is organized as follows.
We introduce the system model in Section~\ref{sec:model}. In Section~\ref{sec:CovProb}, we analyze the \ac{SIR} and network coverage under CEA-ZF and CEU-ZF precoding, also providing simulations that confirm the accuracy of our analysis. We show the numerical results in Section~\ref{sec:NumAnal} to quantify the benefits of CEA-ZF precoding and obtain design insights. We conclude the paper in Section~\ref{sec:conclusions}.

\section{System Model}\label{sec:model}

In this section, we introduce the network topology and propagation model, the CSI available at the \ac{BS},
the conventional CEU-ZF precoder, and the proposed CEA-ZF precoder. The main notations used throughout the paper are summarized in Table~\ref{table:notation}.

\subsection{Network Topology}

\begin{table}
\caption{Notation Summary
} \label{table:notation}
\begin{center}
\renewcommand{\arraystretch}{1.3}
\begin{tabular}{c  p{5.5cm} }
\hline
 {\bf Notation} & {\hspace{2.5cm}}{\bf Definition}
\\
\hline
$\Phi_{\mathrm{b}}$; $\lambda$ & PPP modeling the location of \acp{BS}; \ac{BS} deployment density \\
$N$; $K$ & Number of transmit antennas per \acp{BS}; number of scheduled \acp{UE} per \ac{BS} \\
$\PMT$; $\alpha$ & BS transmit power; path loss exponent \\
$\mathcal{C}_i$; $\mathcal{C}_i^{\mathrm{N}}$ & First-order Voronoi cell for \ac{BS} $i$; cell neighborhood for \ac{BS} $i$ \\
$\mathcal{C}_i^{\mathrm{E}} = \mathcal{C}_i \cup \mathcal{C}_i^{\mathrm{N}}$  & Extended cell for \ac{BS} $i$\\
$r_{iik}$; $r_{\bar{i}ik}$ & Distance between \ac{UE} $k$ in cell $i$ and its serving \ac{BS} $i$ and second closest \ac{BS} $\bar{i}$, respectively \\
$\mathbf{x}_{ijk} \sim \mathcal{CN}(0,\mathbf{I}_N)$ & Small-scale fading between \ac{BS} $i$ and \ac{UE} $k$ in cell $j$\\
$\mathbf{w}_{\mathrm{u},ik}$; $\mathbf{w}_{\mathrm{a},ik}$ & CEU-ZF and CEA-ZF precoding vector, respectively, at \ac{BS} $i$ for its \ac{UE} $k$ \\
$M$; $F$ & Available number of pilots; pilot reuse factor \\
$\mathcal{I}_{\mathrm{p}}$ & Interference during training phase \\
$\mathcal{I}_{\mathrm{u}}$; $\mathcal{I}_{\mathrm{a}}$ & Interference during data transmission phase for CEU-ZF and CEA-ZF, respectively \\
$\tau_{ijk}$  & Standard deviation of CSI error between \ac{BS} $i$ and \ac{UE} $k$ in cell $j$ \\
$\gamma_{ik}$; $\theta$ &  SIR at UE $k$ in cell $i$; SIR decoding threshold \\
$P_{\mathrm{c}}(\theta)$; $\rho_{95}$ & Coverage probability; $95\%$-likely rate \\
\hline
\end{tabular}
\end{center}\vspace{-0.63cm}
\end{table}%

We consider the downlink of a cellular network that consists of randomly deployed BSs, whose location follows a homogeneous \ac{PPP} $\Phi_\mathrm b$ of spatial density $\lambda$ in the Euclidean plane.\footnote{PPPs serve as a good model for the planned deployment of macro cell BSs, as verified by both empirical evidence \cite{TayDhiNov:12} and theoretical analysis \cite{BlaKarKee:13}.} We assume that each \ac{BS} transmits with power $\PMT$ and is equipped with a large number of antennas, $N$.\footnote{We note that power control, which may result in different transmit powers at the BSs, could be used in addition to interference suppression to increase the performance of cell-edge UEs.} Single-antenna \acp{UE} are distributed as an independent homogeneous \ac{PPP} with sufficient high density on the plane, such that each \ac{BS} has at least $K$ candidate \acp{UE} in its cell, i.e., within its coverage area, to transmit to. In light of its higher spectral efficiency, we consider spatial multiplexing at the \acp{BS}, where in each time-frequency RB each \ac{BS} simultaneously serves the $K$ \acp{UE} in its cell with $K\leq N$ \cite{Mar:10}.

We assume that \acp{UE} associate to the \ac{BS} that provides the largest average received power. Due to the homogeneous nature of the network, this results in a distance-based association rule.\footnote{Different association rules apply when transmit power or large-scale fading vary among \acp{BS}, resulting in a weighted Voronoi diagram \cite{YanGerQue:16,SinZhaAnd:15}.} The set of \ac{UE} locations that are associated to \ac{BS} $i$ located at $\mathbf{z}_{i} \in \mathbb{R}^2$ are defined by a classical Voronoi tessellation on the plane, denoted by $\mathcal{V}_{i}^{1}$ and given by \cite{Lee:82,BacGio:15}
\begin{equation}
\mathcal{V}_{i}^{1} = \left\{\mathbf{z} \in \mathbb{R}^2 | \Vert \mathbf{z} - \mathbf{z}_{i}\Vert \leq \Vert \mathbf{z} - \mathbf{z}_k\Vert, ~ \forall ~\mathbf{z}_k \in \Phi_{\mathrm b} \! \setminus \! \mathbf{z}_{i} \right\}.
\end{equation}
We note that the set $\mathcal{V}_{i}^{1}$ contains all locations for which \ac{BS} $i$ is the closest. Such a definition is identical to that of a traditional cell, thus we equivalently denote $\mathcal{V}_i^1$ as $\mathcal{C}_i$.

\begin{figure*}[htp]
  \centering
  \psfrag{CE1}[Bl][Bl][0.65]   {$\mathcal{C}_1$}
  \psfrag{CE2}[Bl][Bl][0.65]   {$\mathcal{C}_2$}
  \psfrag{CE3}[Bl][Bl][0.65]   {$\mathcal{C}_3$}
  \psfrag{CE4}[Bl][Bl][0.65]   {$\mathcal{C}_4$}
  \psfrag{CE5}[Bl][Bl][0.65]   {$\mathcal{C}_5$}
  \psfrag{CE6}[Bl][Bl][0.65]   {$\mathcal{C}_6$}
  \psfrag{CE7}[Bl][Bl][0.65]   {$\mathcal{C}_7$}
  \psfrag{CE8}[Bl][Bl][0.65]   {$\mathcal{C}_8$}
  \psfrag{CE9}[Bl][Bl][0.65]   {$\mathcal{C}_9$}
  \psfrag{CE10}[Bl][Bl][0.65]   {$\mathcal{C}_{10}$}
  \psfrag{CE11}[Bl][Bl][0.65]   {$\mathcal{C}_{11}$}

  \psfrag{V12}[Bl][Bl][0.65]   {$\mathcal{V}_{1,2}^2$}
  \psfrag{V14}[Bl][Bl][0.65]   {$\mathcal{V}_{1,4}^2$}
  \psfrag{V15}[Bl][Bl][0.65]   {$\mathcal{V}_{1,5}^2$}
  \psfrag{V16}[Bl][Bl][0.65]   {$\mathcal{V}_{1,6}^2$}
  \psfrag{V17}[Bl][Bl][0.65]   {$\mathcal{V}_{1,7}^2$}
  \psfrag{V23}[Bl][Bl][0.65]   {$\mathcal{V}_{2,3}^2$}
  \psfrag{V27}[Bl][Bl][0.65]   {$\mathcal{V}_{2,7}^2$}
  \psfrag{V29}[Bl][Bl][0.65]   {$\mathcal{V}_{2,9}^2$}
  \psfrag{V34}[Bl][Bl][0.65]   {$\mathcal{V}_{3,4}^2$}
  \psfrag{V39}[Bl][Bl][0.65]   {$\mathcal{V}_{3,9}^2$}
  \psfrag{V310}[Bl][Bl][0.65]   {$\mathcal{V}_{3,10}^2$}
  \psfrag{V311}[Bl][Bl][0.65]   {$\mathcal{V}_{3,11}^2$}
  \psfrag{V45}[Bl][Bl][0.65]   {$\mathcal{V}_{4,5}^2$}
  \psfrag{V411}[Bl][Bl][0.65]   {$\mathcal{V}_{4,11}^2$}
  \psfrag{V56}[Bl][Bl][0.65]   {$\mathcal{V}_{5,6}^2$}
  \psfrag{V511}[Bl][Bl][0.65]   {$\mathcal{V}_{5,11}^2$}
  \psfrag{V67}[Bl][Bl][0.65]   {$\mathcal{V}_{6,7}^2$}
  \psfrag{V68}[Bl][Bl][0.65]   {$\mathcal{V}_{6,8}^2$}
  \psfrag{V78}[Bl][Bl][0.65]   {$\mathcal{V}_{7,8}^2$}
  \psfrag{V79}[Bl][Bl][0.65]   {$\mathcal{V}_{7,9}^2$}
  \psfrag{V89}[Bl][Bl][0.65]   {$\mathcal{V}_{8,9}^2$}
  \psfrag{V2V11}[Bl][Bl][0.65]   {$\mathcal{V}_{10,11}^2$}
  \psfrag{V910}[Bl][Bl][0.65]   {$\mathcal{V}_{9,10}^2$}

  \subfigure[\label{fig:1a}]{\includegraphics[width=1.0\columnwidth]{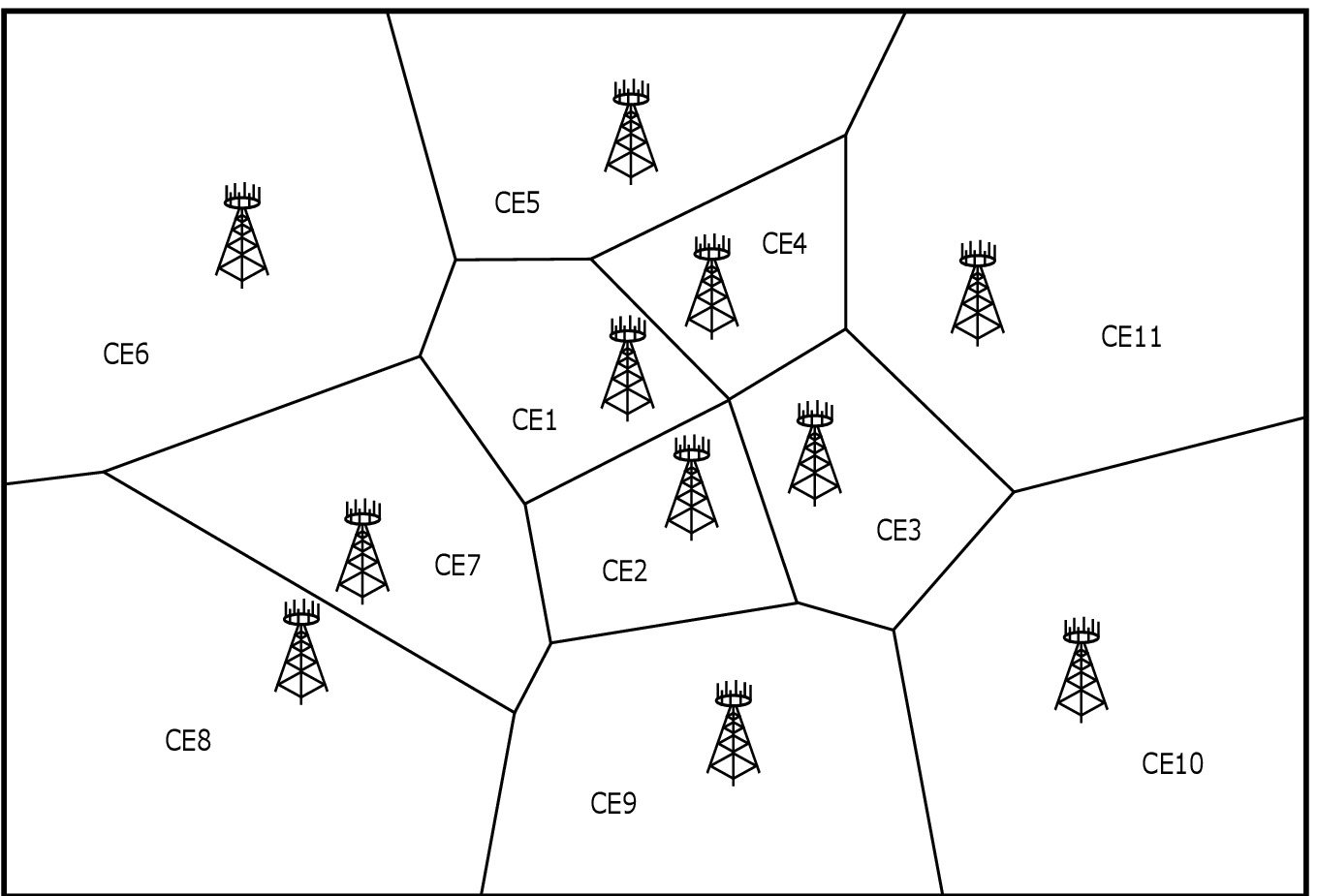}}\quad
  \subfigure[\label{fig:1b}]{\includegraphics[width=1.0\columnwidth]{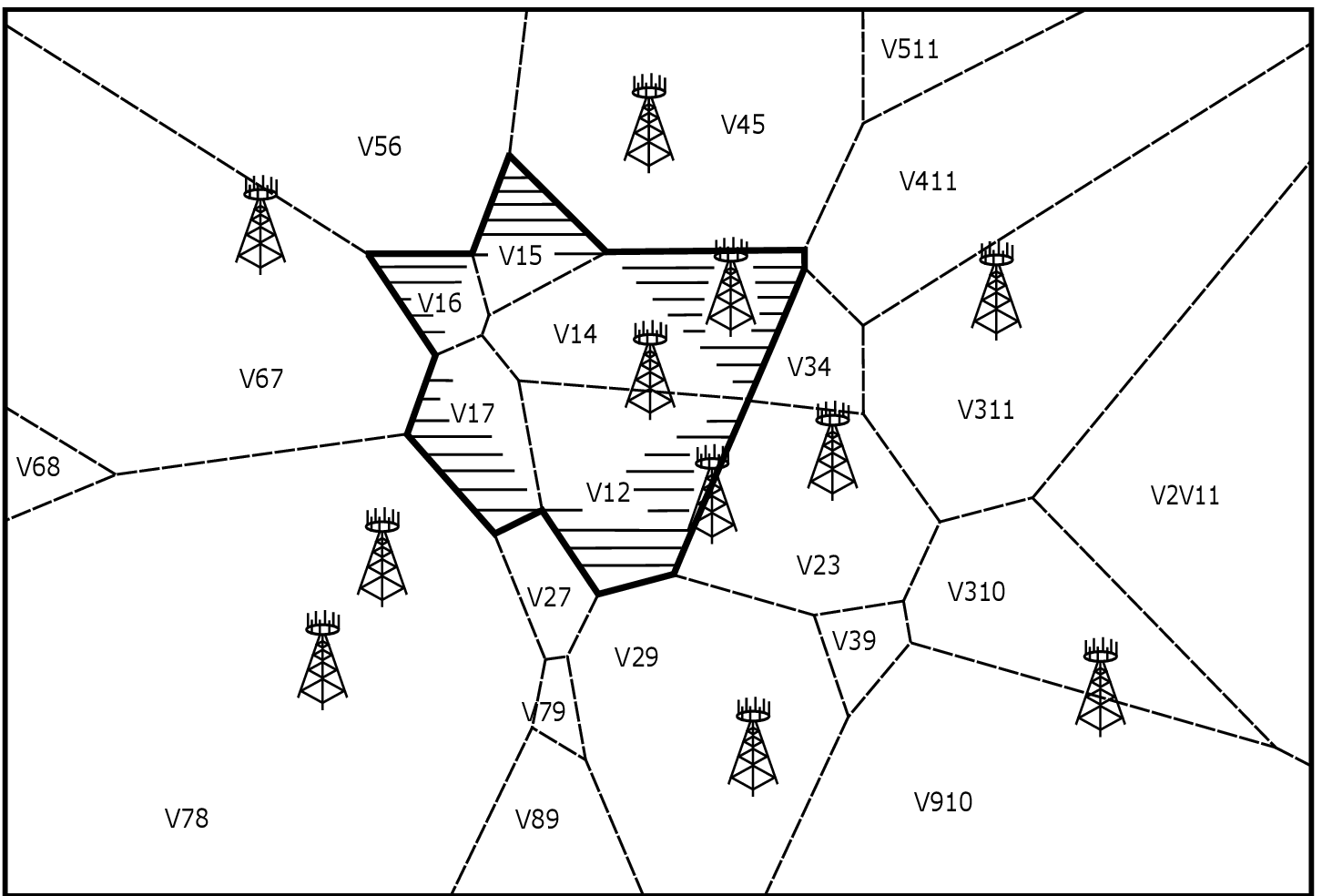}}
  \caption{Examples of (a) first-order and (b) second-order Voronoi tessellation. In (a), solid lines delimit first-order Voronoi cells $\mathcal{C}_i$. In (b), dashed lines delimit second-order Voronoi cells $\mathcal{V}_{i,j}^2$, solid lines delimit the extended cell $\mathcal{C}_{1}^{\mathrm{E}}$, and a shadowed region indicates the cell neighborhood $\mathcal{C}_{1}^\mathrm{N}$.}
  \label{fig:sys_model}
\end{figure*}

In this work, we design a CEA-ZF precoder that controls the interference generated by each BS at the neighboring UEs. In order to identify the neighboring \acp{UE} for each \ac{BS}, we find it useful to generalize the above definition of Voronoi cell to the second order. More precisely, the second-order Voronoi tessellation $\mathcal{V}_{i,j}^{2}$ denotes the set of \ac{UE} locations for which the \acp{BS} in $\mathbf{z}_{i}$ and $\mathbf{z}_{j}$ are the two closest, and it is given by \cite{Lee:82,BacGio:15}
\begin{align}
\mathcal{V}_{i,j}^{2} = &\big\{\mathbf{z} \in \mathbb{R}^2 | \cap_{l \in \{i,j\}}  \{ \Vert \mathbf{z} - \mathbf{z}_{l}\Vert \leq \Vert \mathbf{z} - \mathbf{z}_k\Vert\} \nonumber\\
&    \forall ~\mathbf{z}_k \in \Phi_{\mathrm b} \setminus \{\mathbf{z}_{i},  \mathbf{z}_{j} \} \big\}.
\end{align}

Using the second-order Voronoi tessellation, we can now define the notion of extended cell $\mathcal{C}_i^{\mathrm{E}}$ for \ac{BS} $i$, given by (i) all \acp{UE} for which \ac{BS} $i$ is the closest, and (ii) all \acp{UE} for which \ac{BS} $i$ is the second closest. The extended cell $\mathcal{C}_i^{\mathrm{E}}$ is given by
\begin{align}\label{equ:extended_cell}
\mathcal{C}_i^{\mathrm{E}} = \cup_j \mathcal{V}_{i,j}^{2},\quad  \forall ~\mathbf{z}_j \in \Phi_{\mathrm b} \setminus  \mathbf{z}_{i}.
\end{align}
According to the above definition, each \ac{UE} that lies in $\mathcal{C}_i^{\mathrm{E}}$  sees \ac{BS} $i$ as either its closest or second closest \ac{BS}. The extended cell $\mathcal{C}_i^{\mathrm{E}}$ includes the \acp{UE} located in $\mathcal{C}_i$ that are served by \ac{BS} $i$, as well as the neighboring \acp{UE} which are most vulnerable to interference generated by \ac{BS} $i$. These \acp{UE} constitute the cell neighborhood for \ac{BS} $i$, which we denote by $\mathcal{C}_i^{\mathrm{N}}$ and define as follows
\begin{align}\label{equ:CellNeighbor}
\mathcal{C}_i^{\mathrm{N}} = \mathcal{C}_i^{\mathrm{E}} \setminus \mathcal{C}_{i}.
\end{align}

Fig.~\ref{fig:sys_model} illustrates the concepts of first-order and second-order Voronoi tessellation, cell neighborhood, and extended cell. Fig. \ref{fig:1a} shows a realization of first-order Voronoi tessellation, where each \ac{BS} $i$ covers a cell $\mathcal{C}_{i}$. Fig.~\ref{fig:1b} depicts the corresponding second-order Voronoi tessellation, where each pair of \acp{BS} $(i,j)$ identifies a region $\mathcal{V}_{i,j}^{2}$ (delimited by dashed lines), such that \acp{UE} located in $\mathcal{V}_{i,j}^{2}$ have \ac{BS} $i$ and \ac{BS} $j$ as their closest and second closest, or vice versa. Fig.~\ref{fig:1b} also shows the extended cell $\mathcal{C}_{1}^{\mathrm{E}}$ for \ac{BS} $1$ (delimited by solid lines), which is composed by the first-order cell $\mathcal{C}_{1}$ and by the neighborhood $\mathcal{C}_{1}^\mathrm{N}$ (shadowed region).

In this paper, we propose a CEA-ZF precoding scheme where each \ac{BS} not only spatially multiplexes the associated \acp{UE} in $\mathcal{C}_{i}$, but also suppresses the interference caused at the most vulnerable neighboring \acp{UE} in $\mathcal{C}_i^{\mathrm{N}}$. We note that each \ac{BS} $i$ can easily obtain a list of \acp{UE} in $\mathcal{C}_i^{\mathrm{N}}$ by means of reference signal received power (RSRP) estimation. In fact, downlink RSRP measurements for a list of neighboring \acp{BS} are periodically sent by each \ac{UE} for handover purposes \cite{LopGuvChu:12}.

\subsection{Channel Model and Estimation}

In this network, we model the channels between any pair of antennas as independent and identically distributed (i.i.d.) and quasi-static, i.e., the channel is constant during a sufficiently long coherence block, and varies independently from block to block.\footnote{Note that the results obtained through the machinery of random matrix theory can be modified to model transmit antenna correlation \cite{CouDeb:11, WagCouDeb:12, GerAlnYua:13}.}
Moreover, we assume that each channel is narrowband and affected by two attenuation components, namely small-scale Rayleigh fading,  and large-scale path loss.\footnote{For the sake of tractability, the analysis presented here does not consider shadowing. Note that the results involving large-system approximations can be adjusted to account for the presence of shadowing as in \cite{YanGerQue:16}. Moreover, a generalized gamma approximation can still be used under shadowing, since the channel attenuation at a given \ac{UE} follows a Rayleigh distribution \cite{SinZhaAnd:15}.} As such, the channel matrix from \ac{BS} $i$ to its $K$ associated \acp{UE} can be written as
\begin{align}
{\mathbf{H}_i} = \mathbf{R}_i^{\frac{1}{2}} \mathbf{X}_i
\end{align}
where $\mathbf{R}_i = \text{diag}\{r_{ii1}^{-\alpha}, \cdots, r_{iiK}^{-\alpha}\}$ is the path loss matrix, with $r_{ijk}$ denoting the distance from the \ac{BS} $i$ to \ac{UE} $k$ in cell $j$, i.e., associated with \ac{BS} $j$. The constant $\alpha$ represents the path loss exponent, whereas $\mathbf{X}_i = [\mathbf{x}_{ii1},\cdots,\mathbf{x}_{iiK}]^{\text{H}}$ is the $K \times N$ fading matrix, where $\mathbf{x}_{ijk} \sim \mathcal{CN}(0,\mathbf{I}_N)$ is the channel fading vector between \ac{BS} $i$ and \ac{UE} $k$ in cell $j$. Due to the interference-limited nature of massive MIMO cellular networks,
we neglect the effect of thermal noise \cite{Mar:10}.

In order to simultaneously amplify the desired signal at the intended \acp{UE} and suppress interference at other \acp{UE}, each \ac{BS} requires CSI from all the \acp{UE} it serves.
This {CSI} is obtained during the training phase, where some RBs are used for the transmission of pilot signals. Since the number of pilots, i.e., the number of RBs allocated to the training phase, is limited, these pilots must be reused across cells. Pilot reuse implies that the estimate for the channel between a \ac{BS} and one of its \acp{UE} is contaminated by the channels between the \ac{BS} and \acp{UE} in other cells which share the same pilot \cite{Mar:10,RusPerBuoLar:2013,LuLiSwi:14,MulCotVeh:2014,FerGerQue:16}.

Pilot contamination can be a limiting factor for the performance of massive MIMO. In order to mitigate this phenomenon, non-universal pilot reuse has been proposed, where neighboring cells use different sets of mutually orthogonal pilots \cite{BjoLarDeb:16,BjoLar:15}. Under non-universal pilot reuse, the total set of available pilot sequences is divided into sub-groups, and different sub-groups are assigned to adjacent cells. For a pilot reuse factor $F$, the same sub-group of orthogonal pilot sequences is reused in every $F$ cells.

We denote by $M=\kappa L$ the number of available orthogonal pilots, with $L$ being the number of symbols that can be transmitted within a time-frequency coherence block, and $\kappa$ being the fraction of symbols that are allocated for channel estimation.
For a time-division duplexing (TDD) system with $L=2\times 10^4$ and $\kappa = 5\%$, there would be $M=1000$ orthogonal pilots, and therefore a pilot reuse factor $F=7$ would allow the estimation of 142 \ac{UE} channels per cell \cite{LozHeaAnd:13}.\footnote{In a TDD system, downlink channels can be estimated through uplink pilots thanks to channel reciprocity. This makes the training time proportional to the number of \acp{UE}. A frequency-division duplexing (FDD) system requires a considerably longer training time, proportional to the number of \ac{BS} antennas, and is therefore less suitable for massive MIMO \cite{Mar:10,MarHoc:06}.} As a general rule, a pilot reuse factor $F \geq 3$ is recommended in order to mitigate pilot contamination \cite{BjoLarMar:15}. In this regard, we assume that there are sufficient pilot sequences to support a large enough pilot reuse factor, such that each \ac{BS} can estimate the {CSI} of \acp{UE} in its own cell and in adjacent cells.

In this paper, we express the estimated small-scale fading $\hat{\mathbf{x}}_{ijk}$ between \ac{BS} $i$ and \ac{UE} $k$ in cell $j$ as
\begin{align}
\hat{\mathbf{x}}_{ijk} = \sqrt{1-\tau_{ijk}^2} \mathbf{x}_{ijk} +  \tau_{ijk}  \mathbf{q}_{ijk},
\end{align}
where  $\mathbf{q}_{ijk} \sim \mathcal{CN}\left(\mathbf{0},  \mathbf{I}_N\right)$ is an independent normalized estimation noise and $\tau_{ijk}^2$ is the error variance \cite{MulCouBjo:15, GerCouYua:13, WagCouDeb:12,NosBehAnd:11, WanMur:06}. The parameter $\tau^2_{ijk} \in [0, 1]$ reflects the accuracy of the estimated fading channel $\hat{\mathbf{x}}_{ijk}$, i.e., $\tau^2_{ijk} = 0$ corresponds to perfect CSI while $\tau^2_{ijk} = 1$ corresponds to no CSI at all. By applying MMSE as the channel estimation criterion, the CSI error variance at the typical UE can be written as \cite{hoy:13massive,BjoLarDeb:16}
\begin{align}\label{equ:CSI_ERROR_General}
\tau_{ijk}^2 &= 1 -  \frac{ r_{ijk}^{-\alpha} }{\sum_{l \in  \Phi_{\mathrm{P}}  } r_{ilk}^{-\alpha}}
\nonumber\\
&= \frac{1}{1 + \frac{r_{ijk}^{-\alpha}}{\sum_{l \in  \Phi_{\mathrm{P}} \setminus \{j\} } r_{ilk}^{-\alpha}} }
\end{align}
where $\Phi_{\mathrm{P}}$ indicates the set of \acp{BS} that have their \acp{UE} re-using the same pilot as \ac{UE} $k$ in cell $j$ and are thus generating pilot contamination.
As such, the estimated channel matrix at \ac{BS} $i$ can therefore be written as
\begin{align}\label{equ:estimate_channel}
    \hat{\mathbf{H}}_i = \mathbf{R}_i^\frac{1}{2} \hat{\mathbf{X}}_i.
\end{align}
Despite its simplicity, the above CSI model allows us to investigate the impact of imperfect CSI on the performance of the proposed CEA-ZF precoder.


\subsection{Downlink Transmission}

We now introduce two downlink transmission schemes: (i) the conventional CEU-ZF precoder and (ii) the proposed CEA-ZF precoder.

\subsubsection{Conventional CEU-ZF precoding}

With conventional zero forcing transmission, each \ac{BS} $i$ calculates its precoding matrix as \cite{CouDeb:11, GerEgaYua:12, PeeHocSwi:05}
\begin{align}\label{equ:ZF}
\mathbf{W}_{\mathrm{u},i} = \frac{1}{\sqrt{\zeta_{\mathrm{u},i}}} \cdot \hat{\mathbf{H}}_i^{\mathrm{H}}  \left( \hat{\mathbf{H}}_i \hat{\mathbf{H}}_i^{\mathrm{H}}\right)^{-1},
\end{align}
where $\hat{\mathbf{H}}_i = [\hat{\mathbf{h}}_{ii1}, \cdots, \hat{\mathbf{h}}_{iiK} ]^{\mathrm{H}} \in \mathbb{C}^{K \times N}$ is the estimated channel matrix, and $\zeta_{\mathrm{u},i}$ is a power normalization factor given by
\begin{align}
\zeta_{\mathrm{u},i} = \mathrm{tr}\left[ \left( \hat{\mathbf{H}}_i \hat{\mathbf{H}}_i^{\mathrm{H}} \right)^{-1}\right].
\end{align}
The individual precoding vector from BS $i$ to its UE $k$ is the $k^\mathrm{th}$ column of the precoding matrix.

Note that ZF precoding has been shown to outperform maximum ratio transmission (MRT) in terms of per-cell sum-rate \cite{BjoLarDeb:16}. When the system dimensions make the ZF matrix inversion in (\ref{equ:ZF}) computationally expensive, a simpler truncated polynomial expansion can be employed with similar performance \cite{KamMulBjo:2014}.

\subsubsection{Proposed CEA-ZF precoding}
\begin{figure*}[htp]
  \centering

    {\includegraphics[width=2.0\columnwidth]{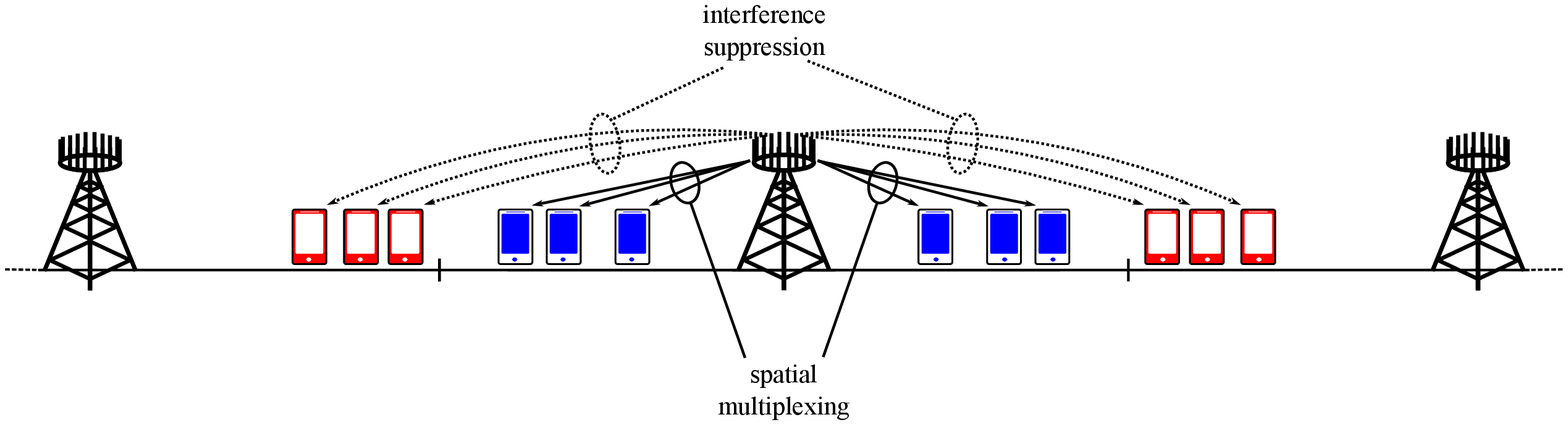}}

  \caption{Illustration of CEA-ZF where a \ac{BS} performs spatial multiplexing for in-cell \acp{UE} and interference suppression for neighboring \acp{UE}.}
  \label{fig:CEA_ZF}
\end{figure*}

Unlike  CEU-ZF precoding, where all spatial dimensions available at each \ac{BS} $i$ are used to multiplex \acp{UE} within cell $\mathcal{C}_i$, the proposed CEA-ZF precoder exploits some spatial dimensions to suppress interference at the most vulnerable \acp{UE}, i.e., those lying in the \ac{BS}'s cell neighborhood $\mathcal{C}_i^{\mathrm{N}}$. Such interference suppression is performed by all \acp{BS} in a distributed manner, and improves the cell-edge performance of the network, as well as the overall data rate and coverage. An illustration of the basic features of CEA-ZF is given in Fig.~\ref{fig:CEA_ZF}, where a multi-antenna \ac{BS} spatially multiplexes its in-cell \acp{UE} while simultaneously suppressing interference at its neighboring \acp{UE}.

For \ac{BS} $i$, we denote by $K'$ the number of \acp{UE} lying in the cell neighborhood, where we omit the subscript $i$ for notational convenience. We note that $K'$ indicates the number of \acp{UE} for which \ac{BS} $i$ is the second closest, i.e., the number of neighboring \acp{UE} for \ac{BS} $i$. The proposed CEA-ZF precoding matrix for \ac{BS} $i$ is then given by
\begin{align}\label{equ:cea_zf}
\mathbf{W}_{\mathrm{a},i} = \frac{1}{\sqrt{\zeta_{\mathrm{a},i}}} \cdot \left\{ \hat{\mathbf{H}}_{i,\mathrm{E}}^{\mathrm{H}} \left(  \hat{\mathbf{H}}_{i,\mathrm{E}} \hat{\mathbf{H}}_{i,\mathrm{E}}^{\mathrm{H}} \right)^{-1} \right\}_{[1:K]},
\end{align}
where $\hat{\mathbf{H}}_{i,\mathrm{E}} = [ \hat{\mathbf{h}}_{ii1}, ... , \hat{\mathbf{h}}_{iiK}, \hat{\mathbf{h}}_{i\bar{i}1}, ..., \hat{\mathbf{h}}_{i\bar{i}K'}]^{\mathrm{H}} \in \mathbb{C}^{(K+K')\times N}$,
with $\{\mathbf{H}\}_{[1:K]}$ indicating the first $K$ columns of the matrix $\mathbf{H}$, and $\hat{\mathbf{h}}_{i\bar{i}l}$ denoting the estimated channel between \ac{BS} $i$ and the $l$-th neighboring \ac{UE}, where the notation $\bar{i}$ indicates that \ac{BS} $i$ is the second closest \ac{BS} for that particular \ac{UE}. The constant $\zeta_{\mathrm{a},i}$  is chosen as an average power normalization factor, given by
\begin{align}
\zeta_{\mathrm{a},i}  = \mathrm{tr}\left[ \left( \left\{ \hat{\mathbf{H}}_{i,\mathrm{E}}^{\mathrm{H}} \left(  \hat{\mathbf{H}}_{i,\mathrm{E}} \hat{\mathbf{H}}_{i,\mathrm{E}}^{\mathrm{H}} \right)^{-1} \right\}_{[1:K]} \right)^2 \right].
\end{align}
We note that the CEA-ZF precoder in \eqref{equ:cea_zf} can be seen as a generalization of the two-cell precoder proposed in \cite{ZakHan:12} to a non-symmetric and non-pairwise scenario.
{\remark{ \textit{The pseudo inverse in \eqref{equ:cea_zf} aims at projecting interference onto the null space spanned by the channels of all UEs in the extended cell $\mathcal{C}_i^{\mathrm{E}}$. In other words, CEA-ZF sacrifices certain degrees of freedom to null the BS interference towards not only in-cell UEs but also out-cell edge UEs. As a result, edge UEs can enjoy lower inter-cell interference, at the expense of fewer degrees of freedom being available at the BSs to provide spatial multiplexing gain. A tradeoff therefore exists between the amount of interference to be suppressed and the remaining degrees of freedom for spatial multiplexing. In fact, the higher the number of out-cell UEs to which a BS suppresses interference, the smaller the number of spatial dimensions for multiplexing gain, and vice versa. Nevertheless, as long as the total number of degrees of freedom is larger than the total number of spatial dimensions needed for UE multiplexing and interference suppression, transmissions from a BS to the served UEs are guaranteed to be interference free under CEA-ZF.} } }

\section{Coverage Analysis}\label{sec:CovProb}

In this section, we analyze the downlink \ac{SIR} coverage of a massive MIMO cellular network with conventional CEU-ZF precoding and the proposed CEA-ZF precoding, and then provide simulations that verify the accuracy of our analysis.

\subsection{Preliminaries}

\subsubsection{SIR at a typical UE}

By applying Slivnyak's theorem to the stationary  \ac{PPP} of \acp{BS}, it is sufficient to evaluate the \ac{SIR} of a typical \ac{UE} at the origin \cite{BacBla:09}. In the following, we denote as \textit{typical} the \ac{UE} $k$ associated with \ac{BS} $i$, with a received signal given by
\begin{align}
y_{ik} \!= & \PMT {\mathbf{h}_{iik}^{\mathrm{H}} \mathbf{w}_{ik} s_{ik}} + \!\!\! \sum_{l=1, l \neq k}^{\KM}  \!\!\! \PMT  {\mathbf{h}_{iik}^{\mathrm{H}} \mathbf{w}_{il} s_{il}}
\nonumber\\
&+  \sum_{j \neq i}^{\infty} \sum_{l=1}^{\KM} \PMT {\mathbf{h}_{jik}^{\mathrm{H}} \mathbf{w}_{jl} s_{jl}}
\end{align}
where $\mathbf{w}_{ik} \in \mathbb{C}^{N \times 1}$ is the normalized precoding vector from the serving \ac{BS} $i$ to the typical \ac{UE}, and $s_{ik}$ is the corresponding unit-power signal, i.e., $\mathbb{E}\left[ \vert s_{ik} \vert^2\right] = 1$. The vector $\mathbf{w}_{ik}$ can take different forms, depending on the precoding scheme employed. The \ac{SIR} at the typical  \ac{UE} can be written as
\begin{align}\label{equ:sinr_ue}
\gamma_{ik} = \frac{ \vert \mathbf{h}_{iik}^{\mathrm{H}} \mathbf{w}_{ik} \vert^2 }{\sum_{l=1,l \neq k}^K \vert \mathbf{h}_{iik}^{\mathrm{H}} \mathbf{w}_{il} \vert^2 + \mathcal{I} },
\end{align}
where the first summation in the denominator represents the intra-cell interference, while $\mathcal{I}$ denotes the aggregate out-of-cell interference. The latter is given by
\begin{align}
\mathcal{I} = \sum_{j \neq i} \frac{g_{jik}}{r_{jik}^{\alpha}},
\end{align}
where $g_{jik}$ is the effective small-scale fading from  interfering \ac{BS} $j$ to \ac{UE} $k$ in cell $i$, given by
\begin{align}\label{equ:effc_fade}
g_{jik} = \sum_{l=1}^K  \left| \mathbf{x}_{jik}^{\mathrm{H}} \mathbf{w}_{jl} \right|^2.
\end{align}
{We note that when the precoding vectors $\{\mathbf{w}_{jl}\}_{l=1}^K$ at BS $j$ are mutually independent and satisfy $\sum_{l=1}^K \Vert \mathbf{w}_{jl} \Vert^2 = 1$, the effective channel fading is distributed as $g_{jik} \sim \Gamma\left(K,1/K\right)$ \cite{GioDhiAnd:14}.}

\subsubsection{Coverage probability}

Since both the received signal strength and the interference at a given UE are governed by a number of stochastic processes, e.g., random spatial distribution of transmitting/receiving nodes and small-scale fading, the resulting SIR is a random variable. In our analysis, the performance metric of interest is the coverage probability, defined as the probability that the received \ac{SIR} $\gamma$ at a generic \ac{UE} is above a given threshold $\theta$, i.e.,
\begin{align}
P_\mathrm{c}(\theta) = \mathbb{P}\left( \gamma \geq \theta \right), \quad \theta > 0.
\end{align}
We note that the coverage probability $P_\mathrm{c}(\theta)$ provides information on the UE \ac{SIR} (and therefore achievable rate) distribution across the network.

\subsubsection{CSI error}

Under sufficient non-universal pilot reuse, a generic \ac{BS} $i$ can estimate the channels of all in-cell \acp{UE} as well as the channels of neighboring \acp{UE}. From \eqref{equ:CSI_ERROR_General}, the {CSI} error variance for an in-cell \ac{UE} and for a neighboring \ac{UE} can be respectively written as
\begin{align}\label{equ:tau_cls}
\tau^2 &= \frac{1}{1+\frac{t^{-\alpha}}{\mathcal{I}_{\mathrm{p}}} },\\ \label{equ:tau_scl}
\bar{\tau}^2 &= \frac{1}{1+\frac{s^{-\alpha}}{\mathcal{I}_{\mathrm{p}}} }
\end{align}
where $t$ and $s$ denote the distance between a typical \ac{UE} and its closest and second closest \ac{BS}, respectively, and $\mathcal{I}_{\mathrm{p}}$ is the pilot interference received during the training phase.

Under reuse factor $F$, clusters of $F$ adjacent cells choose different sub-groups of pilot sequences and do not cause interference, i.e., pilot contamination to each other. Therefore, each \ac{BS} receives pilot contamination only from \acp{UE} lying outside the cluster of $F$ cells, whose mean area can be calculated as $F/\lambda$ \cite{BacBla:09}. This area can be approximated with a circle $B(0,R_{\mathrm{e}})$ of radius $R_e = \sqrt{F/(\lambda \pi)}$ \cite{HeaKouBai:13}, yielding the following mean interference via Campbell's theorem \cite{BacBla:09}
\begin{align}\label{equ:MeanPilotPower}
\mathbb{E}\left[\mathcal{I}_{\mathrm{p}}\right] &= \mathbb{E}\left[ \sum_{x \in \Phi_{\mathrm{P}} \cap B^c(0,R_{\mathrm{e}})} \| x \|^{-\alpha} \right]
\nonumber\\
&= \frac{2 \left( \lambda \pi / F \right)^{\frac{\alpha}{2}}}{\alpha-2},
\end{align}
where $B^c(0,R_{\mathrm{e}})$ denotes the complement set of $B(0,R_{\mathrm{e}})$. By approximating the interfence $\mathcal{I}_{\mathrm{p}}$ with its mean \cite{RalSinAnd:14} and by substituting \eqref{equ:MeanPilotPower} into \eqref{equ:tau_cls} and \eqref{equ:tau_scl}, the {CSI} error can be approximated by a function of $t$ and $s$ as follows:
\begin{align}\label{equ:tau_cls_aprox}
\tau^2 \approx \frac{1}{1+\frac{ \left( \alpha-2 \right) F^{\frac{\alpha}{2}} }{2 \left( \lambda \pi \right)^{\frac{\alpha}{2}} t^{\alpha}} },\\ \label{equ:tau_scl_aprox}
\bar{\tau}^2 \approx \frac{1}{1+\frac{ \left( \alpha-2 \right) F^{\frac{\alpha}{2}} }{2 \left( \lambda \pi \right)^{\frac{\alpha}{2}} s^{\alpha}} }.
\end{align}

\subsection{CEU-ZF Precoding}

We now derive the coverage probability under CEU-ZF precoding.
An approximation of the \ac{SIR} under CEU-ZF can be obtained in the large-system regime as follows.

\begin{lemma}\label{lma:DE_SCP_SIR}
\emph{
Conditioned on the out-of-cell interference $\mathcal{I}_{\mathrm{u}}$ and the intra-cell distance $r_{iil}$, $l \in \{1,\cdots, K\}$,
when $K,N \rightarrow \infty$ with fixed $\beta = K/N < 1$, the \ac{SIR} achieved by CEU-ZF precoding converges almost surely to the following quantity
\begin{align}\label{equ:lsa_scp_sir}
\gamma_{\mathrm{u},ik} - \frac{\left( 1 - \tau_{{iik}}^2\right)(1-\beta) N}{\left(\tau_{iik}^2 r_{iik}^{-\alpha} +  \mathcal{I}_{\mathrm{u}}\right) \left(r_{iik}^{\alpha} + R_k\right)} \rightarrow 0
\end{align}
where $R_k$ is given by
\begin{align} \label{equ:Rk}
R_k = \sum_{l=1,l \neq k}^K r_{iil}^{\alpha}.
\end{align}
}
\end{lemma}
\begin{IEEEproof}
See Appendix~\ref{apx:DE_SCP_SIR}.
\end{IEEEproof}
The accuracy of Lemma~\ref{lma:DE_SCP_SIR} will be verified in Fig.~\ref{fig:del_sinr_vs_n}.

Deriving the coverage probability requires knowledge of the distribution of $R_k$, which is the sum of $(K-1)$ i.i.d. random variables (r.v.s) $r_{lii}^\alpha$. 
The distance $r_{lii}$ is a r.v. that follows a Rayleigh distribution $f_{\mathrm{c}}(r)$, given by \cite{hae:12}
\begin{align}\label{equ:cls_dist}
f_{\mathrm{c}}(r) = 2\pi\lambda r e^{-\lambda\pi r^2}.
\end{align}
It can then be shown that $r_{lii}^{\alpha}$ follows a Weibull distribution with shape and scale parameters $2/\alpha$ and $(\lambda \pi)^{-\frac{\alpha}{2}}$, respectively \cite{Bil:08}.
As such, the distribution of $R_k$
can be approximated by a generalized Gamma distribution as follows \cite{FilYac:06}.
\begin{assumption}\label{def:GenGam}
\emph{The probability density function (pdf) $f_{R_k}(r)$ and cumulative density function (CDF) $F_{R_k}(r)$ of the r.v. $R_k$ can be approximated as follows
\begin{align}\label{equ:pdf_GenGam}
f_{R_k}(r) &\approx \frac{\eta \mu^\mu r^{\eta \mu - 1} }{\Omega^\mu \Gamma(\mu)} \exp\left( - \frac{\mu r^\eta}{\Omega}\right),\\ \label{equ:cdf_GenGam}
F_{R_k}(r) &\approx \frac{1}{\Gamma\left( \mu \right)} \Gamma\left( \mu, \frac{\mu r^\eta}{\Omega} \right),
\end{align}
where $\Gamma(s,x)=\int_0^x t^{s-1} e^{-t} dt$ is the lower incomplete gamma function, and $\Omega = \mathbb{E}[R_k^\eta]$ is a scale parameter, given by
\begin{align}\label{equ:Omega_Expression}
\Omega = \left[ \frac{\mu^{\frac{1}{\eta}} \Gamma\left( \mu \right) \mathbb{E}[R_k]}{\Gamma\left( \mu + \frac{1}{\eta} \right)} \right]^\eta
\end{align}
while $\mu$ and $\eta$ are solutions of the following equations
\begin{align}
\frac{\Gamma^2\left( \mu + \frac{1}{\eta}\right)}{\Gamma\left( \mu \right) \Gamma \left( \mu + \frac{2}{\eta}\right) - \Gamma^2\left( \mu + \frac{1}{\eta}\right)} = \frac{\mathbb{E}^2[R_k]}{\mathbb{E}[R_k^2] - \mathbb{E}^2[R_k]},\\
\frac{\Gamma^2\left( \mu + \frac{2}{\eta}\right)}{\Gamma\left( \mu \right) \Gamma \left( \mu + \frac{4}{\eta}\right) - \Gamma^2\left( \mu + \frac{2}{\eta}\right)} = \frac{\mathbb{E}^2[R_k^2]}{\mathbb{E}[R_k^4] - \mathbb{E}^2[R_k^2]}.
\end{align}
The quantities $\mathbb{E}[R_k]$, $\mathbb{E}[R_k^2]$, and $\mathbb{E}[R_k^4]$ are the first, second, and fourth moment of the r.v. $R_k$, respectively, and can be calculated as
\begin{align}\label{equ:ERk}
&\mathbb{E}[R_k] = \frac{K-1}{\left( \lambda \pi \right)^{\frac{\alpha}{2}}}   \Gamma\!\left( 1 \!+\! \frac{\alpha}{2}\right),\\ \label{equ:Rk2}
&\mathbb{E}[R_k^2] = \frac{K-1}{\left( \lambda \pi \right)^{{\alpha}}}   \left[ \Gamma\!\left( 1 \!+\! {\alpha}\right) \!+\! \left( K \!-\! 2 \right)  \Gamma^2\!\left( 1 \!+\! \frac{\alpha}{2}\right) \right],\\
&\mathbb{E}[R_k^4] = \frac{K\!-\!1}{\left( \lambda \pi \right)^{{2 \alpha}}} \left[ \left(K\!-\!2\right)\left(K\!-\!3\right)\left(K\!-\!4\right)\Gamma^4 \! \left( 1 \!+\! \frac{\alpha}{2}\right)
\right.
\nonumber\\
&+\! 3\!\left(K\!-\!2\right) \Gamma^2\!\left( 1 \!+\! {\alpha}\right)  \!+\! 4 \left(K\!-\!2\right)\Gamma\!\left( 1 \!+\! \frac{\alpha}{2}\right)\!\Gamma\!\left(\! 1 \!+\! \frac{3\alpha}{2}\right)
\nonumber\\
&+\! \left.\Gamma\!\left( 1 \!+\! 2 {\alpha}\right) \!+\! 6 \left(K\!-\!2\right)\left(K\!-\!3\right) \Gamma\!\left( 1 \!+\! {\alpha}\right)\Gamma^2\!\left( 1 \!+\! \frac{\alpha}{2}\right)\right].
\end{align}
}
\end{assumption}
The accuracy of the approximation in Assumption~\ref{def:GenGam} will be verified in Fig.~\ref{fig:weibull_approx}.

By using the approximated distribution of $R_k$, we can now obtain the coverage probability of a massive MIMO cellular network under CEU-ZF.
\begin{theorem}\label{thm:OP_SCP}
\emph{The coverage probability of a massive MIMO cellular network under CEU-ZF precoding can be approximated as
\begin{align}\label{equ:CovProb_SCP}
&\mathbb{P}\left( \gamma_{\mathrm{u},ik} \geq \theta \right)
\nonumber\\
 &\approx \frac{1}{\Gamma  \left( \mu \right)} \! \int\limits_0^\infty \! \Gamma \! \left(\! \mu, \frac{\mu}{\Omega}\! \left[ \frac{ \left( 1 \!-\! \tau^2\right)\! \left( 1 \!-\! \beta \right) N  r^\alpha }{\theta \left( \tau^2 \!+\! \frac{2\pi \lambda r^{2}}{\alpha - 2}\right)} - r^\alpha \right]^\eta\right) \! f_{\mathrm{c}}(r) dr
\end{align}
where $\tau^2$ is given in \eqref{equ:tau_cls_aprox}, and $f_{\mathrm{c}}(r)$ is given by \eqref{equ:cls_dist}.}
\end{theorem}
\begin{IEEEproof}
See Appendix~\ref{apx:OP_SCP}.
\end{IEEEproof}

We note that although coverage probability of a multi-user MIMO cellular network with conventional CEU-ZF precoding has also been derived in \cite{DhiKouAnd:13,LiZhaLet:14,GioDhiAnd:14}, the result in \eqref{equ:CovProb_SCP} provides an approximation that involves only one integration and is therefore easier to be evaluated. The accuracy of this approximation will be verified in Fig. \ref{fig:StoGeo_SCP_CBF}.

\subsection{CEA-ZF Precoding}

We now derive the coverage probability under the proposed CEA-ZF precoder. An approximation of the SIR under CEA-ZF can be obtained in the large-system regime as follows.

\begin{lemma}\label{lma:DE_CBF_SIR}
\emph{Conditioned on the out-of-cell interference $\mathcal{I}_{\mathrm{a}}$, the intra-cell distance $r_{iil}$ with $l \in \{1,\cdots, K\}$, the distance $r_{\bar{i}ik}$ between the typical \ac{UE} and its second closest \ac{BS}, and the standard deviation $\tau_{\bar{i}ik}$ of the corresponding CSI error, when $K,N \rightarrow \infty$ with fixed $\beta = K/N < 1$ and fixed $\beta'=K'/N < 1$, the \ac{SIR} of CEA-ZF converges almost surely to a quantity given by
\begin{align}\label{equ:lsa_cbf_sir}
\gamma_{\mathrm{a},ki} - \frac{\left( 1 - \tau_{{iik}}^2\right)(1-\beta-\beta') N}{\left(\tau_{iik}^2 r_{iik}^{-\alpha} + \tau_{\bar{i}ik}^2 r_{\bar{i}ik}^{-\alpha} +  \mathcal{I}_{\mathrm{a}}\right)\left(r_{iik}^{\alpha} + R_k \right) } \rightarrow 0.
\end{align}
}
\end{lemma}
\begin{IEEEproof}
See Appendix \ref{apx:DE_CBF_SIR}.
\end{IEEEproof}
The accuracy of Lemma~\ref{lma:DE_CBF_SIR} will be verified in Fig.~\ref{fig:del_sinr_vs_n}.

Using the above results, we are now able to derive the coverage probability under CEA-ZF precoding.
\begin{theorem}\label{thm:OP_CBF}
\emph{The coverage probability of a massive MIMO cellular network under CEA-ZF can be approximated as
\begin{align}
&\mathbb{P}\left( \gamma_{\mathrm{a},ik} \geq \theta \right)
\nonumber\\ \label{equ:OP_GenKm}
&\approx  \int\limits_0^\infty \int\limits_t^\infty\! \mathbb{E}_{K'}\!\!\! \left[ \! \Gamma \! \left(\! \mu, \frac{\mu}{\Omega} \left[ \frac{ \left( 1 \!-\! \tau^2\right)\left( 1 \!-\!  \beta - \beta' \right) N s^\alpha }{ \theta \left( \bar{\tau}^2 + \tau^2 \frac{s^{\alpha}}{t^{\alpha}}  \!+\! \frac{2\pi \lambda s^{2}}{\alpha - 2} \right)} - t^\alpha \right]^\eta\right)\!\right]
 \nonumber\\
 & \qquad \qquad \times \frac{f_{\mathrm{s}|\mathrm{c}}(s,t) f_{\mathrm{c}}(t)}{\Gamma  \left( \mu \right)} \,  ds dt
\end{align}
where  $\tau^2$ and $\bar{\tau}^2$ are given in \eqref{equ:tau_cls_aprox} and \eqref{equ:tau_scl_aprox}, respectively, and
\begin{align}\label{equ:scl_dist}
f_{\mathrm{s}|\mathrm{c}}(s,t) = 2 \pi \lambda s e^{-\lambda \pi \left( s^2 - t^2 \right)}.
\end{align}
}
\end{theorem}
\begin{IEEEproof}
See Appendix~\ref{apx:CP_CBF}.
\end{IEEEproof}

The result in \eqref{equ:OP_GenKm} involves an expectation on the number $K'$ of neighboring \acp{UE}. While the number $K$ of \acp{UE} associated to each BS is constant and depends on the scheduling process, $K'$ is a r.v. whose distribution is generally unknown. In the following, we provide a more compact approximation for the coverage probability under CEA-ZF by approximating $K'$ with its mean value, derived as follows.

\begin{proposition}\label{prop:second_voronoi}
\emph{The expected number of neighboring \acp{UE} for each \ac{BS}, $K'$, satisfies
\begin{align}
\mathbb{E}\left[ K' \right]  = K.
\end{align}
}
\end{proposition}
\begin{IEEEproof}
See Appendix \ref{apx:second_voronoi}.
\end{IEEEproof}

\begin{corollary}\label{cor:CovCEAZF}
\emph{By replacing $K'$ with its mean $K$, the coverage probability of a massive MIMO cellular network under CEA-ZF can be further approximated as
\begin{align}
&\mathbb{P}\left( \gamma_{\mathrm{a},ik} \geq \theta \right) \label{equ:OP_ApxKm}
\nonumber\\
 &\approx \int\limits_0^\infty \int\limits_t^\infty \!  \Gamma \! \left(\! \mu, \frac{\mu}{\Omega} \left[ \frac{ \left( 1 \!-\! \tau^2\right)\left( 1 \!-\!  2 \beta  \right) N s^\alpha }{ \theta \left( \bar{\tau}^2 + \tau^2 \frac{s^{\alpha}}{t^{\alpha}}  \!+\! \frac{2\pi \lambda s^{2}}{\alpha - 2} \right)} - t^\alpha \right]^\eta\right)
 \nonumber\\
 & \qquad \qquad \times \frac{f_{\mathrm{s}|\mathrm{c}}(s,t) f_{\mathrm{c}}(t)}{\Gamma  \left( \mu \right)} \,  ds dt.
\end{align}
}
\end{corollary}
\begin{IEEEproof}
The result follows from \eqref{equ:OP_GenKm} by approximating $\beta' = K' / N$ with its mean value $\beta = K / N$.
\end{IEEEproof}
The accuracy of Corollary~\ref{cor:CovCEAZF} will be verified in Fig.~\ref{fig:StoGeo_NumKm}.

\subsection{Asymptotic Results}\label{sec:asym_result}
Equations \eqref{equ:CovProb_SCP} and \eqref{equ:OP_ApxKm} quantify how some of the key features of a cellular network, i.e., deployment strategy, interference, and impairments in the channel estimation phase, affect the coverage probability under CEA-ZF and CEU-ZF precoding. Based on these results, simple asymptotic expressions for the coverage probability can be obtained when perfect CSI is assumed available at the massive MIMO BSs.
\begin{corollary}\label{cor:special_case}
\emph{Under perfect CSI, i.e., $\tau = \bar{\tau} = 0$, when $\alpha =4$ and $\beta = \frac{K}{N} \ll 1$, the network coverage probabilities under CEU-ZF and CEA-ZF can be respectively approximated by the following quantities
\begin{align}\label{equ:CovProb_CEUZF_Asym}
&\mathbb{P}\left( \gamma_{\mathrm{u},ik} \geq \theta \right) \approx 1 - {2\beta \theta} = 1 - \frac{2 K \theta}{N}, \\ \label{equ:CovProb_CEAZF_Asym}
&\mathbb{P}\left( \gamma_{\mathrm{a},ik} \geq \theta \right) \approx 1 - \left( {2 \beta \theta}\right)^2  = 1 - \frac{4 K^2 \theta^2}{N^2}.
\end{align}
}
\end{corollary}
\begin{IEEEproof}
See Appendix~\ref{apx:special_case}.
\end{IEEEproof}
The two following observations can be readily made from (41) and (42).

\textbf{Observation 1:} \textit{In the asymptotic regime, CEA-ZF achieves higher coverage probability than CEU-ZF only when $\beta < \frac{1}{2 \theta}$, or equivalently, $N > 2 \theta K$. This observation indicates that CEA-ZF requires a sufficiently large number of BS antennas to attain good performance.}

\textbf{Observation 2:} \textit{For a large number of antennas, $N$, the outage probabilities under CEA-ZF, $1 - \mathbb{P}\left( \gamma_{\mathrm{a},ik} \geq \theta \right)$, and under CEU-ZF, $1 - \mathbb{P}\left( \gamma_{\mathrm{u},ik} \geq \theta \right)$,
decay as $\frac{1}{N^2}$ and $\frac{1}{N}$, respectively. These scaling laws prove CEA-ZF more effective for coverage enhancement in massive MIMO cellular networks.}

We note that although the approximations \eqref{equ:CovProb_CEUZF_Asym} and \eqref{equ:CovProb_CEAZF_Asym} only hold asymptotically, a similar behavior can be observed under imperfect CSI and for a finite number of BS antennas, as we will show in Section~\ref{sec:NumAnal}.

\subsection{Analysis Validation}
We now show simulation results that confirm the accuracy of our analytical framework. Unless differently specified, we use the following parameters for path loss exponent, \ac{BS} density, and number of \ac{UE}, respectively: $\alpha=4$, $\lambda = 10^{-6} \mathrm{m}^{-2} = 1 / \mathrm{km}^2$, and $K=10$.

In Fig.~\ref{fig:del_sinr_vs_n}, we depict the downlink \ac{SIR} achieved by a typical \ac{UE} of a massive MIMO cellular network as a function of the number of \ac{BS} antennas $N$, under different precoding schemes and transmit CSI errors. The figure shows a negligible difference between simulations and analytical results, which confirms the accuracy of Lemma~\ref{lma:DE_SCP_SIR} and Lemma~\ref{lma:DE_CBF_SIR}. We also note that the \ac{SIR} values obtained for ZF precoding are consistent with the ones obtained in \cite{hoy:13massive}.

\begin{figure}[t!]
\begin{center}
{
    \psfrag{ANTENNANUMBERN}[Bl][Bl][0.75]   {Number of \ac{BS} antennas, $N$}
	\psfrag{SIRNU}[Bl][Bl][0.85]   {SIR, $\gamma$}
	\psfrag{TAUEQU00}[Bl][Bl][0.70]   {$\tau^2=0.0$}
    \psfrag{TAUBQU00}[Bl][Bl][0.70]   {$\bar{\tau}^2=0.0$}
	\psfrag{TAUEQU01}[Bl][Bl][0.70]   {$\tau^2=0.1$}
    \psfrag{TAUBQU02}[Bl][Bl][0.70]   {$\bar{\tau}^2=0.2$}
	\psfrag{SIMULATIONSCEAZFZ}[Bl][Bl][0.70]   {Simulations: CEA-ZF}
    \psfrag{SIMULATIONSZF}[Bl][Bl][0.70]   {Simulations: CEU-ZF}
	\psfrag{ANALYSISCEAZF}[Bl][Bl][0.70]   {Analysis: CEA-ZF}
	\psfrag{ANALYSISZF}[Bl][Bl][0.70]   {Analysis: CEU-ZF}
    \includegraphics[width=1.0\columnwidth]{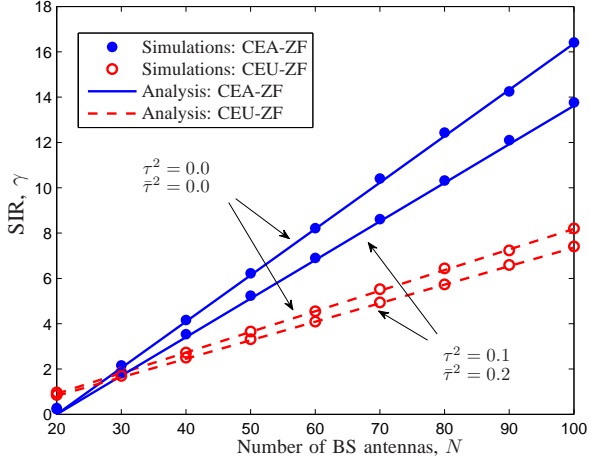}
	}
\end{center}
\caption{Downlink SIR for CEU-ZF and CEA-ZF versus number of \ac{BS} antennas, for scenarios with and without {CSI} error.
}
		\label{fig:del_sinr_vs_n}
\end{figure}

In Fig. \ref{fig:weibull_approx}, we compare the simulated CDF of the r.v. $R_k$ in \eqref{equ:Rk} to the generalized gamma approximation proposed in Assumption~\ref{def:GenGam}, for various values of the number of scheduled \ac{UE} per cell, $K$. The figure shows a close match for all values of $K$, which confirms the accuracy of Assumption~\ref{def:GenGam}.

\begin{figure}[t!]
\begin{center}
{
    \psfrag{x}[Bl][Bl][0.85]   {$x$}
	\psfrag{CDFOFSIPRK}[Bl][Bl][0.85]   {CDF of $R_k$, $F_{R_k}(x)$}
	\psfrag{SIMULATIONSS}[Bl][Bl][0.70]   {Simulations}
    \psfrag{ANALYSIS}[Bl][Bl][0.70]   {Analysis}
    \psfrag{KEQU10}[Bl][Bl][0.70]   {$K=10$}
    \psfrag{KEQU30}[Bl][Bl][0.70]   {$K=30$}
    \psfrag{KEQU50}[Bl][Bl][0.70]   {$K=50$}
	\includegraphics[width=1.0\columnwidth]{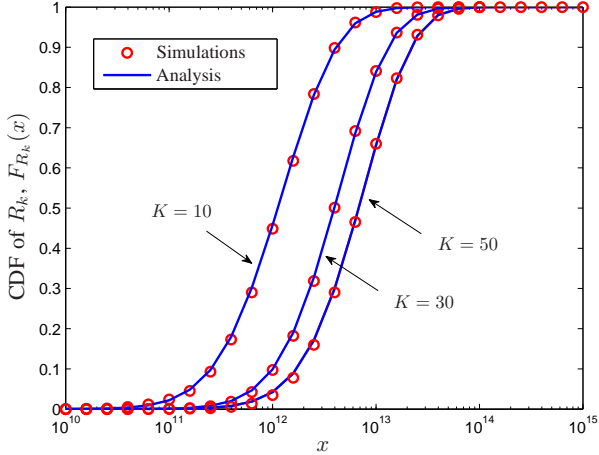}
	}
\end{center}
\caption{CDF of the r.v. $R_k$ in \eqref{equ:Rk} and proposed approximation in Assumption~\ref{def:GenGam}, for \ac{BS} density $\lambda = 10^{-6}$ and different numbers of \ac{UE} per cell, $K$.}
		\label{fig:weibull_approx}
\end{figure}

Fig.~\ref{fig:StoGeo_SCP_CBF} compares the simulated coverage probability to the analytical results derived in Theorem \ref{thm:OP_SCP} and Theorem~\ref{thm:OP_CBF}. The coverage probability is plotted versus the \ac{SIR} threshold at the typical \ac{UE}. The figure shows that analytical results and simulations fairly well match and follow the same trend, thus confirming the accuracy of the theorems.

\begin{figure}[t!]
\begin{center}
{
    \psfrag{SIRTHRESHOLD}[Bl][Bl][0.85]   {SIR threshold, $\theta$}
	\psfrag{COVERAGEPROBAB}[Bl][Bl][0.85]   {Coverage probability, $P_{\mathrm{c}}(\theta)$}
	\psfrag{SIMULATIONSCEAZFS}[Bl][Bl][0.70]   {Simulations: CEA-ZF}
    \psfrag{ANALYSISCEAZF}[Bl][Bl][0.70]   {Analysis: CEA-ZF}
    \psfrag{SIMULATIONSZF}[Bl][Bl][0.70]   {Simulations: CEU-ZF}
    \psfrag{ANALYSISZF}[Bl][Bl][0.70]   {Analysis: CEU-ZF}
	\includegraphics[width=1.0\columnwidth]{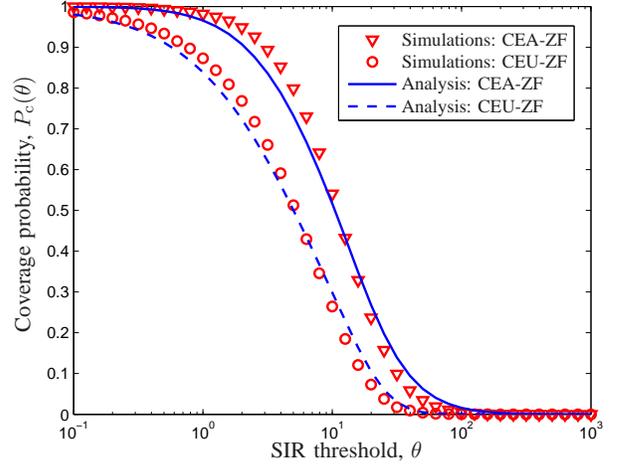}
	}
\end{center}
\caption{Coverage probability under CEU-ZF and CEA-ZF precoding: analysis versus simulations.}
\label{fig:StoGeo_SCP_CBF}
\end{figure}

Fig.~\ref{fig:StoGeo_NumKm} compares the coverage probability derived analytically in Theorem \ref{thm:OP_CBF} and its approximation in Corollary~\ref{cor:CovCEAZF}. It can be seen that the two results match well under different values of $K$, hence confirming the accuracy of Corollary~\ref{cor:CovCEAZF}.

\begin{figure}[t!]
\begin{center}
{
    \psfrag{SIRTHRESHOLD}[Bl][Bl][0.85]   {SIR threshold, $\theta$}
	\psfrag{COVERAGEPROBABILITY}[Bl][Bl][0.85]   {Coverage probability, $P_{\mathrm{c}}(\theta)$}
	\psfrag{ANALYTICAL}[Bl][Bl][0.70]   {Analysis \eqref{equ:OP_GenKm}}
    \psfrag{APPROXIMATIONS}[Bl][Bl][0.70]   {Approx. \eqref{equ:OP_ApxKm}}
    \psfrag{KEQU5}[Bl][Bl][0.70]   {$K = 5$}
    \psfrag{KEQU10}[Bl][Bl][0.70]   {$K = 10$}
    \psfrag{KEQU15}[Bl][Bl][0.70]   {$K = 15$}
	\includegraphics[width=1.0\columnwidth]{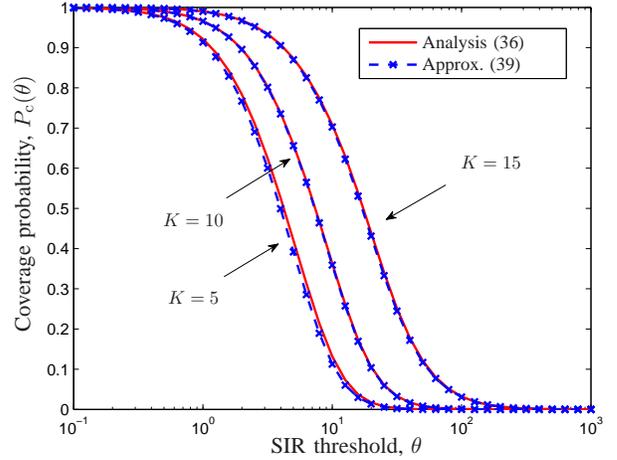}
	}
\end{center}
\caption{Coverage probability under CEA-ZF precoding (Theorem~\ref{thm:OP_CBF}) and its approximation (Corollary~\ref{cor:CovCEAZF}).
}
		\label{fig:StoGeo_NumKm}
\end{figure}


\section{Numerical Results and Discussion}\label{sec:NumAnal}

In this section, we provide numerical results to show the performance gain attained by the proposed CEA-ZF precoder, and we discuss how data rate and coverage are affected by the number of scheduled \acp{UE} and  the channel estimation accuracy.

\subsection{Numerical Results}

Unless otherwise stated, the following system parameters will be used: \ac{BS} deployment density $\lambda = 10^{-6} \mathrm{m}^{-2}$, i.e.,  $\lambda = 1 \mathrm{BS} / \mathrm{km}^2$, number of scheduled \acp{UE} per cell $K=20$, path loss exponent $\alpha=3.8$, and pilot reuse factor $F=7$. In this paper, the data rate for an SIR $\gamma$ is calculated as $\mathcal{R} = K \mathbb{E}[\log_2(1+\gamma)]$, and it does not account for the fraction of time spent for training, since this depends on the pilot allocation scheme employed. We note that neglecting the training time does not affect the fairness of our performance comparison between different precoding schemes. In the following, the performance of the proposed CEA-ZF precoder is compared to the one of CEU-ZF, as well as to two multi-cell MMSE precoders, i.e., the M-MMSE precoder \cite{LiBjoLar:15} and the MC-MMSE precoder \cite{JosAshMar:11}.

Fig.~\ref{fig:CovPro_vs_N} depicts the coverage probability under M-MMSE, MC-MMSE, CEU-ZF, and CEA-ZF precoding as a function of the number of \ac{BS} antennas $N$, for two different \ac{SIR} thresholds $\theta$. The following can be observed: (i) a switching point exists, i.e., CEA-ZF outperforms CEU-ZF precoding only if the number of BS antennas exceeds a certain value, (ii) as the number of BS antennas grows, the coverage probability under CEA-ZF converges to one faster than that under CEU-ZF, and (iii) CEA-ZF attains the best coverage probability when the SIR threshold is low, while MC-MMSE and M-MMSE outperform CEA-ZF in high-SIR regime. We note that observations (i) and (ii) are consistent with the ones we made in Section~\ref{sec:asym_result} in an asymptotic regime with perfect CSI, and observation (iii) comes from the fact that both M-MMSE and MC-MMSE have an intrinsic regularization factor to balance between signal gain and interference, whereas CEA-ZF is dedicated to suppress strong interference at the cell edge and thus is more suitable for low-SIR scenarios.

\begin{figure}[t!]
\begin{center}
{\psfrag{ANTENNANUMBERN}[Bl][Bl][0.85]   {Number of \ac{BS} antennas, $N$}
	\psfrag{COVERAGEPROBAB}[Bl][Bl][0.85]   {Coverage probability, $P_{\mathrm{c}}(\theta)$}

	\psfrag{CEAZFSFEQUE0dB}[Bl][Bl][0.70]   {CEA-ZF, ~~~~\,$\theta = 0 \text{dB}$}
    \psfrag{CEUZFSFEQUE0}[Bl][Bl][0.70]   {CEU-ZF, ~~~~\,$\theta = 0 \text{dB}$}
    \psfrag{EMMSESFEQUEQUEQ0dBs}[Bl][Bl][0.70]   {M-MMSE, ~~~\!$\theta = 0 \text{dB}$}
    \psfrag{JMMSESFEQUE0dB}[Bl][Bl][0.70]   {MC-MMSE, ~\!$\theta = 0 \text{dB}$}
    \psfrag{CEAZFSFEQUE15dB}[Bl][Bl][0.70]   {CEA-ZF, ~~~~\,$\theta = 15 \text{dB}$}
    \psfrag{CEUZFSFEQUE15}[Bl][Bl][0.70]   {CEU-ZF, ~~~~\,$\theta = 15 \text{dB}$}
    \psfrag{EMMSESFEQUEQU15dB}[Bl][Bl][0.70]   {M-MMSE, ~~~\!$\theta = 15 \text{dB}$}
    \psfrag{JMMSESFEQUE15dB}[Bl][Bl][0.70]   {MC-MMSE, ~\!$\theta = 15 \text{dB}$}    
    
		\includegraphics[width=1.0\columnwidth]{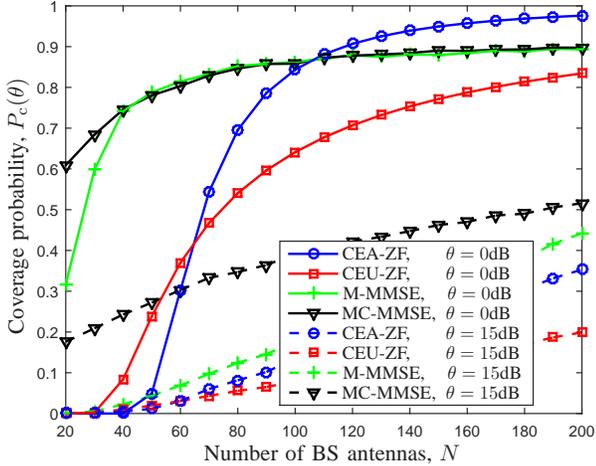}}
\end{center}
\caption{Coverage probability versus number of \ac{BS} antennas for $K=20$ scheduled \acp{UE} per cell, under M-MMSE, MC-MMSE, CEU-ZF, and CEA-ZF precoding.}
		\label{fig:CovPro_vs_N}
\end{figure}

In Fig.~\ref{fig:FivthCovPro_vs_N}, we compare the $95\%$-likely rate under the four precoding schemes. The $95\%$-likely rate (denoted by $\rho_{95}$) is defined as the rate achievable by at least $95\%$ of the \acp{UE} in the network, and it can be regarded as the worst rate any scheduled \ac{UE} may expect to receive when located at the cell edge \cite{AndBuzCho:14,Mar:10}. While the $95\%$-likely rates of both CEU-ZF and CEA-ZF precoding benefit from a larger number of \ac{BS} antennas $N$, the proposed CEA-ZF precoder achieves a significantly larger $95\%$-likely rate compared to conventional CEU-ZF, and the gain increases as $N$ grows. We note that M-MMSE and MC-MMSE can also provide improved 95\%-likely rates over CEU-ZF, as both of them implicitly suppress interference towards cell-edge UEs by balancing between signal gain and interference. Nevertheless, CEA-ZF achieves the best performance in terms of 95\%-likely rate, because it employs fewer spatial dimensions to mitigate inter-cell interference, leaving more degrees of freedom for multiplexing gain.

\begin{figure}[t!]
\begin{center}
{\psfrag{ANTENNANUMBERN}[Bl][Bl][0.85]   {Number of \ac{BS} antennas, $N$}
	\psfrag{95PERCENTRATERHO95}[Bl][Bl][0.85]   {$95\%$-likely rate, $\rho_{95}$}
    
    \psfrag{CEAZFSKEQUE10}[Bl][Bl][0.70]   {CEA-ZF, ~~~~$K=10$}
    \psfrag{SCPKEQUE10}[Bl][Bl][0.70]   {CEU-ZF, ~~~~$K=10$}
    \psfrag{EMMSEKEQUEQEQUE10}[Bl][Bl][0.70]   {M-MMSE,  ~~$K=10$}
    \psfrag{JMMSEKEQUE10}[Bl][Bl][0.70]   {MC-MMSE, $K=10$}

    \psfrag{CEAZFSKEQUE20}[Bl][Bl][0.70]   {CEA-ZF, ~~~~$K=20$}
    \psfrag{SCPKEQUE20}[Bl][Bl][0.70]   {CEU-ZF, ~~~~$K=20$}
    \psfrag{EMMSEKEQUEQUE20}[Bl][Bl][0.70]   {M-MMSE, ~~$K=20$}
    \psfrag{JMMSEKEQUE20}[Bl][Bl][0.70]   {MC-MMSE, $K=20$}

    \raisebox{-1.0\height}{\includegraphics[width=1.0\columnwidth]{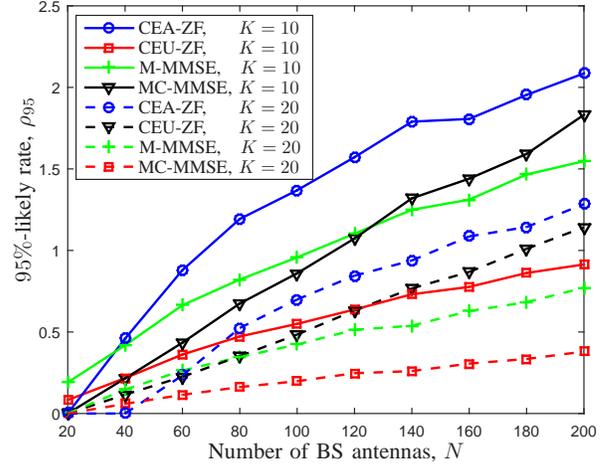}}
    }
\end{center}
\caption{$95\%$-likely rate versus number of BS antennas, with M-MMSE, MC-MMSE, CEU-ZF, and CEA-ZF precoding}
		\label{fig:FivthCovPro_vs_N}
\end{figure}

Fig.~\ref{fig:SmErRt_vs_K} compares the sum-rate per cell under CEU-ZF and CEA-ZF as a function of the number of scheduled \acp{UE} per cell, $K$. An optimal value of $K$ that maximizes the sum-rate exists for both CEU-ZF and CEA-ZF precoding, due to a tradeoff between simultaneously serving more \acp{UE} and having fewer spatial dimensions available for interference suppression. Fewer \acp{UE} should be scheduled under CEA-ZF, thus leaving more spatial dimensions for cell-edge interference suppression, and achieving a higher sum-rate compared to conventional CEU-ZF precoding.

\begin{figure}[t!]
\begin{center}
{\psfrag{NUMBEROFUESK}[Bl][Bl][0.85]   {Number of scheduled \acp{UE} per cell, $K$}
	\psfrag{SUMERGODICRATE}[Bl][Bl][0.85]   {Sum-rate per cell, $\mathcal{R}$}
    \psfrag{SCPNEQUE100}[Bl][Bl][0.70]   {CEU-ZF, $N=100$}
    \psfrag{CEAZFSNEQUE100s}[Bl][Bl][0.70]   {CEA-ZF, $N=100$}
    \psfrag{SCPNEQUE150}[Bl][Bl][0.70]   {CEU-ZF, $N=150$}
    \psfrag{CEAZFSNEQUE150}[Bl][Bl][0.70]   {CEA-ZF, $N=150$}
	\includegraphics[width=1.0\columnwidth]{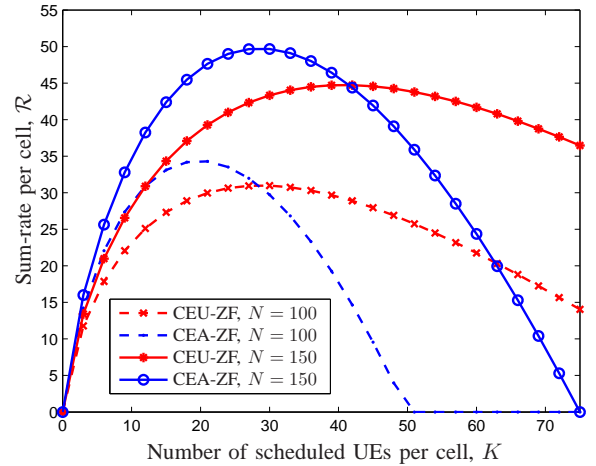}}
\end{center}
\caption{Sum ergodic rate versus number of \acp{UE} per cell, for different numbers of \ac{BS} antennas, $N$.}
		\label{fig:SmErRt_vs_K}
\end{figure}

We now study the impact of the channel estimation error on coverage and edge rates.
To this end, we vary the {CSI} error variance $\tau^2$ and $\bar{\tau}^2$ at in-cell \acp{UE} and neighboring \acp{UE}, respectively, while keeping their ratio constant as $\mathbb{E}[\bar{\tau}^2]/\mathbb{E}[\tau^2]=1.8$. In the following we set the \ac{SIR} threshold as $\theta = 0 ~\mathrm{dB}$ and number of \ac{BS} antennas as $N=100$.

Fig.~\ref{fig:CovProb_vs_tau} shows the coverage probability as a function of the {CSI} error for various values of scheduled \acp{UE} per cell. Although the presence of a {CSI} error degrades the coverage probability of both CEU-ZF and CEA-ZF precoding, it can be seen that CEA-ZF significantly outperforms conventional CEU-ZF for low-to-moderate values of the {CSI} error. Under a large {CSI} error, CEA-ZF still performs as well as CEU-ZF as long as the number of scheduled \acp{UE} per cell is controlled, e.g., $K=10$ or $K=20$ in the figure.

\begin{figure}[t!]
\begin{center}
{\psfrag{COVERAGEPROBAB}[Bl][Bl][0.85]   {Coverage probability, $P_{\mathrm{c}}(\theta)$}
	\psfrag{CSIERRORTAU}[Bl][Bl][0.85]   {CSI error variance, $\tau^2$}
    \psfrag{CEAZFSKEQUE10s}[Bl][Bl][0.70]   {CEA-ZF, $K=10$}
    \psfrag{SCPKEQUE10}[Bl][Bl][0.70]   {CEU-ZF, $K=10$}
    \psfrag{CEAZFSKEQUE20}[Bl][Bl][0.70]   {CEA-ZF, $K=20$}
    \psfrag{SCPKEQUE20}[Bl][Bl][0.70]   {CEU-ZF, $K=20$}
	\includegraphics[width=1.0\columnwidth]{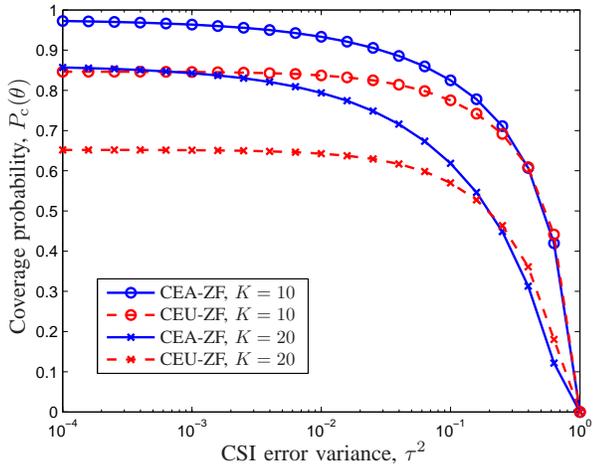}}
\end{center}
\caption{Coverage probability plotted as a function of CSI error variance.}
\label{fig:CovProb_vs_tau}
\end{figure}

Fig.~\ref{fig:FivthRt_vs_tau} depicts the $95\%$-likely rate as a function of the {CSI} error variance for CEU-ZF and CEA-ZF precoding. Once again, CEA-ZF significantly outperforms conventional CEU-ZF for low-to-moderate values of the {CSI} error, while the $95\%$-likely rates of both precoders degrade and achieve similar values under very poor {CSI} quality, i.e., large values of $\tau^2$ and $\bar{\tau}^2$.

\begin{figure}[t!]
\begin{center}
{\psfrag{95PERCENTRHO95}[Bl][Bl][0.85]   {$95\%$-likely rate, $\rho_{95}$}
	\psfrag{CSIERRORTAU}[Bl][Bl][0.85]   {CSI error variance, $\tau^2$}
    \psfrag{CEAZFSKEQUE10s}[Bl][Bl][0.70]   {CEA-ZF, $K=10$}
    \psfrag{SCPKEQUE10}[Bl][Bl][0.70]   {CEU-ZF, $K=10$}
    \psfrag{CEAZFSKEQUE20}[Bl][Bl][0.70]   {CEA-ZF, $K=20$}
    \psfrag{SCPKEQUE20}[Bl][Bl][0.70]   {CEU-ZF, $K=20$}
    \includegraphics[width=1.0\columnwidth]{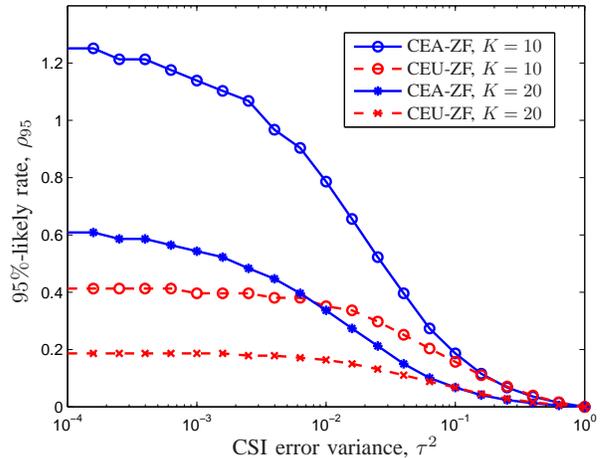}}
\end{center}
\caption{$95\%$-likely rate plotted as a function of CSI error variance.}
\label{fig:FivthRt_vs_tau}
\end{figure}

\subsection{Discussion}\label{SubSec:Discussion}

The main takeaways provided by our analytical framework are outlined as follows.

\subsubsection{Performance gain}

The proposed CEA-ZF precoder outperforms conventional CEU-ZF precoding from several perspectives. CEA-ZF provides better coverage than CEU-ZF, especially in the massive MIMO regime, i.e., when \acp{BS} are equipped with a large number of antennas, $N$. While CEA-ZF can attain high coverage probability with reasonable values of $N$, a significantly larger number of antennas is required to achieve the same coverage under CEU-ZF precoding. The proposed CEA-ZF precoder also achieves a larger sum-rate per cell, and a significantly larger $95\%$-likely rate. The latter is especially important, being the worst data rate that any scheduled \ac{UE} can expect to receive.

\subsubsection{\ac{UE} scheduling}

The aggregate sum-rate per cell is sensitive to the number of \acp{UE} $K$ simultaneously served through spatial multiplexing, for both CEU-ZF and CEA-ZF precoding. It is therefore important to schedule the right number of \acp{UE} for transmission as a function of the number of \ac{BS} antennas, $N$. The proposed CEA-ZF precoder is slightly more sensitive to variations of $K$. The rationale behind such phenomenon is that when $N$ just merely exceeds $K$, sacrificing extra spatial dimensions to suppress interference at neighboring \acp{UE} cannot compensate for the loss in power gain experienced by the in-cell \acp{UE}. On the other hand, when $K \ll N$, not enough \acp{UE} are being multiplexed, and some of the available degrees of freedom are being ``wasted'' in power gain, while they could instead yield higher multiplexing gain \cite{BjoLarMar:15}. We note that multiplexing gain enhances the rate linearly, whereas power gain only enhances the rate logarithmically, i.e., at a lower pace.

\subsubsection{Channel estimation error}

As expected, pilot contamination can negatively affect the achievable data rates by degrading the quality of the {CSI} available at the \acp{BS}. In the presence of very large channel estimation errors, the performance of CEA-ZF precoding degrades and converges to the one of conventional CEU-ZF precoding. In fact, the cell-edge suppression mechanism employed by CEA-ZF relies on the accuracy of the measured channels, and the promised gains in terms of coverage and $95\%$-likely rate cannot be achieved. It is therefore desirable to control the amount of pilot contamination received during the channel estimation phase, for example by designing appropriate pilot allocation schemes.

\section{Conclusions}\label{sec:conclusions}

In this paper, we proposed the CEA-ZF precoder, which exploits the excess spatial degrees of freedom available at massive MIMO \acp{BS} to suppress inter-cell interference at the most vulnerable \acp{UE} in the network. Unlike joint processing techniques, CEA-ZF can be implemented in a distributed fashion. Moreover, CEA-ZF specifically targets those neighboring \acp{UE} close to the \ac{BS} coverage area, thus requiring fewer spatial dimensions to mitigate inter-cell interference, and leaving more dimensions for intra-cell spatial multiplexing.

In order to model practical deployments, we analyzed the performance of CEA-ZF and conventional CEU-ZF precoding in a random asymmetric cellular network. We showed that a higher per-cell data rate and a better network coverage can be guaranteed by the CEA-ZF precoder. More importantly, the $95\%$-likely rate, namely the minimum data rate that any \ac{UE} can expect to achieve, is significantly improved. The latter is of particular interest, given the ambitious edge rate requirements set for 5G, which aims for an uninterrupted high data rate user experience. While this paper focused on the downlink of cellular networks, CEA-ZF can also be employed as an uplink receive filter to remove the interference generated by \acp{UE} in nearby cells. Similar gains are expected in the uplink setting, although this should be verified in future work.
Besides, the proposed CEA-ZF could be combined with suitable power control schemes to further enhance the cell-edge rates.
Additionally, this work modeled the BS deployment with a PPP, which may tend to overestimate the cell edge effects. Evaluating the gains in a more realistic scenario, e.g., by introducing a minimum BS inter-site distance, is also left as future work.

Our study also quantified the impact of imperfect CSI, confirming the importance of controlling the amount of pilot contamination during the channel estimation phase. While our emphasis was on the \ac{SIR} distribution across the network, the data rate may also be affected by the amount of resources dedicated to pilot signals. An inherent tradeoff exists between mitigating pilot contamination and reusing pilot resources. To this end, fractional pilot reuse (FPR) has been proposed in \cite{AtzArnDeb:15,ZhuWanQia:16}, where cell-center \acp{UE} of neighboring cells reuse the same pilots, while cell-edge \acp{UE} employ non-universal pilot reuse. Analyzing the performance of FPR in randomly deployed networks is regarded as an interesting research direction.

\begin{appendix}

\subsection{Proof of Lemma~\ref{lma:DE_SCP_SIR}}\label{apx:DE_SCP_SIR}

Conditioned on the out of cell interference $\mathcal{I}_{\mathrm{u}}$, we substitute the expression of $\mathbf{w}_{{\mathrm{u}},ik}$ in \eqref{equ:ZF} into  \eqref{equ:sinr_ue}, then as $K,N \rightarrow \infty$ with $\beta = K/N < 1$, the \ac{SIR} $\gamma$ converges to the following \cite[Theorem 14.3]{CouDeb:11}
\begin{align}\label{equ:DetEquv_BSDL}
\gamma_{\mathrm{u},ik}  -  \frac{1 - \tau_{iik}^2}{ \Upsilon \cdot r_{iik}^{-\alpha} \tau_{iik}^2  + \Psi \cdot \mathcal{I}_{\mathrm{u}} }  \rightarrow 0, \,\,\, a.s.
\end{align}
where $\Upsilon$ and $\Psi$ are given by
\begin{align}\label{equ:upsilon}
\Upsilon &= \frac{1}{\phi} \frac{1}{N} \mathrm{tr}\mathbf{R}_i^{-1},\\ \label{equ:cap_psi}
\Psi &= \frac{\psi}{\frac{N}{K} \phi^2 - \psi } \frac{1}{K} \frac{1}{N} \mathrm{tr}\mathbf{R}_i^{-1}
\end{align}
and $\psi$ and $\phi$ are given, respectively, by
\begin{align}\label{equ:psi}
\psi &= \frac{1}{N} \mathrm{tr}\left( \mathbf{I}_N + \frac{K}{N} \frac{1}{\phi} \mathbf{I}_N \right)^{-2}
= \frac{1}{\left( 1 + \frac{\beta}{\phi} \right)^2},\\ \label{equ:phi}
\phi &= \frac{1}{N} \mathrm{tr}\left[ \left( 1 + \frac{\beta}{\phi}  \right) \mathbf{I}_N \right]^{-1}
= \frac{1}{1+\frac{\beta}{\phi}}.
\end{align}
By solving \eqref{equ:psi} and \eqref{equ:phi}, we obtain $\phi = 1 - \beta$ and $\psi = \phi^2$.
By  substituting  $\psi$ and $\phi$ in \eqref{equ:upsilon} and \eqref{equ:cap_psi}, respectively, the following holds
\begin{align} \label{equ:upsilon_psi}
\Upsilon = \Psi = \frac{1}{N} \frac{1}{1-\beta}\sum_{l=1}^K r_{iil}^\alpha.
\end{align}
Lemma~\ref{lma:DE_SCP_SIR} then follows by substituting \eqref{equ:upsilon_psi} into \eqref{equ:DetEquv_BSDL}.

\subsection{Proof of Theorem~\ref{thm:OP_SCP}}\label{apx:OP_SCP}

Consider a typical \ac{UE} $k$ in cell $i$ and located at the origin. We denote the distance between the typical \ac{UE} and its serving \ac{BS} $i$ as $r_{iik}=t$. As such, we can approximate the out of cell interference $\mathcal{I}_{\mathrm{u}}$ by its mean, which can be computed as
\begin{align}\label{equ:AveIso}
\mathbb{E}\left[ \mathcal{I}_{\mathrm{u}} \right] &= \mathbb{E}\left[ \sum_{x \in \Phi_{\mathrm{b}} \setminus i } \frac{g_{xik}}{\| x \|^{\alpha}}  \right]
\nonumber\\
& \stackrel{(a)}{=} \int_{t}^\infty r^{-\alpha} r dr = \frac{2 \pi \lambda t^{-(\alpha-2)}}{\alpha-2}
\end{align}
where $(\text{a})$ is obtained by taking expectation of the effective fading $g_{xik}$ and then using Campbell's theorem \cite[Theorem 1.4.3]{BacBla:09}.
By substituting \eqref{equ:AveIso} into \eqref{equ:lsa_scp_sir}, the conditional \ac{SIR} received at the typical \ac{UE} can be approximated as
\begin{align}
{\gamma}_{\mathrm{u},ik} \approx \frac{\left( 1 - \tau_{{iik}}^2\right)(1-\beta) N}{\left(\tau_{iik}^2 t^{-\alpha} +  \frac{2 \pi \lambda t^{-(\alpha-2)}}{\alpha-2}\right) \left(t^{\alpha} + R_k\right)}.
\end{align}
The coverage probability can then be calculated as
\begin{align}\label{equ:cond_cov_scp}
&\mathbb{P}\left({\gamma}_{\mathrm{u},ik}  \geq \theta \right)
\nonumber\\
&\approx
\mathbb{E}\left[ \mathbb{P} \left( \left. R_k \leq \frac{\left( 1 - \tau_{{iik}}^2\right)(1-\beta) N t^\alpha}{\left(\tau_{iik}^2 +  \frac{2 \pi \lambda t^{2}}{\alpha-2}\right) \theta} - t^\alpha \right\vert r_{iik} = t  \right) \right].
\end{align}
Since the pdf of $t$ has been given in \eqref{equ:cls_dist}, Theorem~\ref{thm:OP_SCP} then follows from \eqref{equ:cond_cov_scp} by deconditioning $t$.

\subsection{Proof of Lemma~\ref{lma:DE_CBF_SIR}}\label{apx:DE_CBF_SIR}
We start with the \ac{MMSE} version of the CEA-ZF precoder, which includes a regularization term $ \rho \mathbf{I}_N$, given by
\begin{align}
\tilde{\mathbf{w}}_{\mathrm{a},ik} = \frac{1}{\sqrt{\tilde{\zeta}_{\mathrm{a},i}}}\! \left( \rho N \mathbf{I}_N \!+\!\! \sum_{l=1}^K \hat{\mathbf{h}}_{iil} \hat{\mathbf{h}}_{iil}^{\mathrm{H}} \!+\!\! \sum_{l=1}^{K'} \hat{\mathbf{h}}_{i\bar{i}l} \hat{\mathbf{h}}_{i\bar{i}l}^{\mathrm{H}} \right)^{-1} \!\!\!\!\!\! \hat{\mathbf{h}}_{iik}
\end{align}
where $\tilde{\zeta}_{\mathrm{a},i}$ satisfies that $\sum_{k=1}^K \left\Vert \tilde{\mathbf{w}}_{\mathrm{a},ik} \right\Vert^2 = 1$.
The \ac{SIR} at the typical \ac{UE} can be written as
\begin{align}\label{equ:CAZF_SIR}
\gamma_{\mathrm{a},ik} = \frac{\left\vert \mathbf{h}_{iik}^{\mathrm{H}} \tilde{\mathbf{w}}_{\mathrm{a},ik} \right\vert^2}{\sum_{l \neq k} \left\vert \mathbf{h}^{\mathrm{H}}_{iik} \tilde{\mathbf{w}}_{\mathrm{a},il} \right\vert^2 \!+\! \sum_{l=1}^{K} \left\vert \mathbf{h}^{\mathrm{H}}_{\bar{i}ik} \tilde{\mathbf{w}}_{\mathrm{a},\bar{i}l} \right\vert^2 \!+\! {I}_{\mathrm{a}} }.
\end{align}

Substituting \eqref{equ:estimate_channel} into the numerator of \eqref{equ:CAZF_SIR}, and using the matrix inversion lemma and the rank-1 permutation lemma \cite[Theorem 3.9]{CouDeb:11}, the received signal power converges to the following limit in the large-system regime
\begin{align}
\left\vert {\mathbf{h}}_{iik}^{\mathrm{H}} \tilde{\mathbf{w}}_{\mathrm{a},ik} \right\vert^2 -  \frac{(1-\tau_{iik}^2)}{\zeta_{\mathrm{a},i}}\frac{\left( r_{iik}^{-\alpha} \Lambda_i \right)^2}{\left( 1 + r_{iik}^{-\alpha} \Lambda_i \right)^2} \rightarrow 0, \quad \text{as}~~ N \rightarrow \infty
\end{align}
where $\Lambda_i$ is the solution of the following fix point equation \cite[Theorem 13]{HanTse:01}
\begin{align}
\Lambda_i = \frac{1}{\rho + \frac{1}{N} \sum_{l=1}^K \frac{r_{iil}^{-\alpha}}{1 + \Lambda_i r_{iil}^{-\alpha}} + \frac{1}{N} \sum_{l=1}^{K'} \frac{r_{i\bar{i}l}^{-\alpha}}{1 + \Lambda_i r_{i\bar{i}l}^{-\alpha}} }.
\end{align}
We next deal with the first two summations in the denominator of \eqref{equ:CAZF_SIR}, which are the intra-cell interference from the serving \ac{BS} and the inter-cell interference from the second closest interfering  \ac{BS}.
Similarly, by using the rank-1 permutation again, the large-system limit for these interference reads as
\begin{align}
&\sum_{l \neq k} \left\vert \mathbf{h}^{\mathrm{H}}_{iik} \tilde{\mathbf{w}}_{\mathrm{a},il} \right\vert^2
\nonumber\\
&= \sum_{l} \frac{ \frac{1}{ \zeta_{\mathrm{a},i} } \frac{r_{iil}^{-\alpha}}{N} \left( - \frac{\partial \Lambda_i}{\partial \rho}\right) }{\left( 1 + r_{iil}^{-\alpha} \Lambda_i \right)^2} \left[ \frac{\left( 1 - \tau_{iik}^2 \right) r_{iik}^{-\alpha} }{\left( 1 + r_{iik}^{-\alpha} \Lambda_i \right)^2} + \tau_{iik}^2 r_{iik}^{-\alpha} \right],
\end{align}
and
\begin{align}
&\sum_{l } \left\vert \mathbf{h}^{\mathrm{H}}_{\bar{i}ik} \tilde{\mathbf{w}}_{\mathrm{a},\bar{i}l} \right\vert^2
\nonumber\\
&= \sum_{l} \frac{ \frac{1}{ \zeta_{\mathrm{a},\bar{i}} } \frac{r_{\bar{i}\bar{i}l}^{-\alpha}}{ N } \left( - \frac{\partial \Lambda_{\bar{i}}}{\partial \rho}\right) }{\left( 1 + r_{\bar{i}\bar{i}l}^{-\alpha} \Lambda_{\bar{i}} \right)^2} \left[ \frac{\left( 1 - \tau_{\bar{i}ik}^2 \right) r_{\bar{i}ik}^{-\alpha} }{\left( 1 + r_{\bar{i}ik}^{-\alpha} \Lambda_{\bar{i}} \right)^2} + \tau_{\bar{i}ik}^2 r_{\bar{i}ik}^{-\alpha} \right],
\end{align}
respectively. For the power normalization factor $\tilde{\zeta}_{\mathrm{a},i}$ and $\tilde{\zeta}_{\mathrm{a},\bar{i}}$, the deterministic equivalence under large-system regime can be derived in a similar way as
\begin{align}
\tilde{\zeta}_{\mathrm{a},i} \rightarrow  \frac{1}{N} \sum_{l=1}^K \frac{r_{iil}^{-\alpha}
\left(- \frac{\partial \Lambda_i}{\partial \rho}\right)}{\left( 1 \!+\! r_{iil}^{-\alpha} \Lambda_i  \right)^2}, \quad
\tilde{\zeta}_{\mathrm{a},\bar{i}} &\rightarrow  \frac{1}{N} \sum_{l=1}^K \frac{r_{\bar{i}\bar{i}l}^{-\alpha}
\left( - \frac{\partial \Lambda_{\bar{i}}}{\partial \rho} \right)}{\left( 1 \!+\! r_{\bar{i}\bar{i}l}^{-\alpha} \Lambda_{\bar{i}}  \right)^2}.
\end{align}
As such, we have the deterministic equivalence of SIR being
\begin{align}\label{equ:DetEqv_CBF_Apdx}
\gamma_{\mathrm{a},ik} = \frac{\left( 1 - \tau_{iik}^2 \right) \frac{\left( r_{iik}^{-\alpha} \Lambda_i \right)^2}{\left( 1 + r_{iik}^{-\alpha} \Lambda_i  \right)^2} \left\vert \frac{1}{N} \sum_{l=1}^K \frac{r_{iil}^{-\alpha} \left( - \frac{\partial \Lambda_i}{\partial \rho}\right) }{\left( 1 \!+\! r_{iil}^{-\alpha} \Lambda_i  \right)^2} \right\vert^{-1}  }{ \frac{\left( 1 - \tau_{iik}^2 \right) r_{iik}^{-\alpha} }{\left( 1 + r_{iik}^{-\alpha} \Lambda_i \right)^2} \!+\! \tau_{iik}^2 r_{iik}^{-\alpha} \!+\! \frac{\left( 1 - \tau_{\bar{i}ik}^2 \right) r_{\bar{i}ik}^{-\alpha} }{\left( 1 + r_{\bar{i}ik}^{-\alpha} \Lambda_{\bar{i}} \right)^2} \!+\! \tau_{\bar{i}ik}^2 r_{\bar{i}ik}^{-\alpha} \!+\! {I}_{\mathrm{a}} }.
\end{align}
Finally, by letting $\rho \rightarrow 0$, each term in \eqref{equ:DetEqv_CBF_Apdx} that contains $\Lambda_i$ respectively converges to
\begin{align}\label{equ:lma2_term1}
&\frac{\left( r_{iik}^{-\alpha} \Lambda_i \right)^2}{\left( 1 + r_{iik}^{-\alpha} \Lambda_i  \right)^2} = \frac{\left( r_{iik}^{-\alpha} \rho \Lambda_i \right)^2}{\left( \rho + r_{iik}^{-\alpha} \rho \Lambda_i  \right)^2} \rightarrow 1,\\\label{equ:lma2_term2}
& \frac{ \left( 1 - \tau_{iik}^2 \right)  r_{iik}^{-\alpha} }{\left( 1 + r_{iik}^{-\alpha} \Lambda_i  \right)^2} = \frac{ \left( 1 - \tau_{iik}^2 \right) \rho^2 r_{iik}^{-\alpha} }{\left( \rho + \rho r_{iik}^{-\alpha} \Lambda_i  \right)^2} \rightarrow 0,\\ \label{equ:lma2_term3}
&  \left\vert \frac{1}{N} \sum_{l=1}^K \frac{r_{iil}^{-\alpha} \left( - \frac{\partial \Lambda_i}{\partial \rho} \right)}{\left( 1 \!+\! r_{iil}^{-\alpha} \Lambda_i  \right)^2} \right\vert^{-1} \rightarrow \frac{\left(1 - \beta - \beta' \right)N}{ \sum_{l=1}^K r_{iil}^\alpha }.
\end{align}
Lemma~\ref{lma:DE_CBF_SIR} then follows by substituting \eqref{equ:lma2_term1}, \eqref{equ:lma2_term2}, and \eqref{equ:lma2_term3} into \eqref{equ:DetEqv_CBF_Apdx}.

\subsection{Proof of Theorem~\ref{thm:OP_CBF}}\label{apx:CP_CBF}

We consider a typical \ac{UE} $k$ of \ac{BS} $i$ that locates at the origin, and denote the distance between the \ac{UE} and its associated \ac{BS} as $r_{iik}=t$ and the distance from the \ac{UE} to its second closest \ac{BS} as $r_{\bar{i}ik} = s$. As such, we can approximate the out of cell interference $\mathcal{I}_{\mathrm{a}}$ by its mean based on Campbell's theorem \cite{BacBla:09}, given as follows
\begin{align}\label{equ:AveIco}
\mathbb{E}\left[ \mathcal{I}_{\mathrm{a}} \right] = \mathbb{E}\left[ \sum_{x \in \Phi_{\mathrm{b}} \setminus \{i, \bar{i} \}  } \frac{g_{xik}}{\Vert x \Vert^{\alpha}} \right]
= \frac{2\pi \lambda s^{-(\alpha - 2)}}{\alpha - 2}.
\end{align}
By substituting \eqref{equ:AveIco} into \eqref{equ:lsa_cbf_sir}, the conditional \ac{SIR} received at the typical \ac{UE} can be approximated as
\begin{align}
\gamma_{\mathrm{a},ik} \approx \frac{(1-\tau_{iik}^2)(1-\beta - \beta')N }{\left( \tau_{iik}^2 t^{-\alpha} + \tau_{\bar{i}ik}^2 s^{-\alpha} + \frac{2 \pi \lambda s^{-(\alpha-2)}}{\alpha-2} \right) \left( t^\alpha + R_k \right) }.
\end{align}
The coverage probability can then be approximated as
\begin{align}\label{equ:cond_cov_cbf}
&\mathbb{P}\left( \gamma_{\mathrm{a},ik} \geq \theta \right)
\nonumber\\
&\approx \mathbb{P}\left( R_k \leq \frac{\left( 1 - \tau_{iik}^2 \right) \left( 1 - \beta - \beta' \right) N \theta^{-1} }{\tau_{iik}^2 t^{-\alpha} + \tau_{\bar{i}ik}^2 s^{-\alpha} + \frac{2 \pi \lambda s^{-(\alpha-2)} }{\alpha-2} } - t^\alpha \right).
\end{align}
Theorem~\ref{thm:OP_CBF} then follows from \eqref{equ:cond_cov_cbf} by deconditioning on $t$ and $s$, with their pdf given in \eqref{equ:cls_dist} and \eqref{equ:scl_dist}, respectively.

\subsection{Proof of Proposition~\ref{prop:second_voronoi}}\label{apx:second_voronoi}

Since the second-order Voronoi cells form a tessellation on $\mathbb{R}^2$ \cite{BacBla:09}, the cell centers form a stationary point process $\Phi_{\mathrm{b}}'$ with density $\lambda'$.
Without loss of generality, we consider the center of the second-order Voronoi cell $\mathcal{V}_{x,y}^2$ as a typical point in $\Phi_{\mathrm{b}}'$ located at the origin.
We further denote the two \acp{BS} $x$ and $y$ that identify $\mathcal{V}_{x,y}^2$ as out-neighbors for the typical point 0.
In this sense, under  Palm probability  $\mathbf{P'}^0$, the set of out-neighbors for the origin 0 is formally defined as
\begin{align}
h^+(\omega) = \left\{ y \in \Phi_{\mathrm{b}}: 0 \in \mathcal{V}_{y, z}^2, \quad z \in \Phi_{\mathrm{b}} \setminus \{y\} \right\}.
\end{align}
By leveraging compatibility, we can define the out-neighbors of any point $X \in \Phi_{\mathrm{b}}'$ as follows
\begin{align}
H^+\left( X \right) = X + h^+\left( T_{X} \right)
\end{align}
where $T_X$ is the shift operator defined in \cite{DalVer:07}.

On the other hand, if BS $x$ in $\Phi_\mathrm{b}$ is regarded as typical, we can define in-neighbors for this typical point to be centers of all the second-order Voronoi cells constructed by BS $x$.
Specifically, under Palm probability $\mathbf{P}^0$, the set of in-neighbors for the origin 0 and  any $Y \in \Phi_{\mathrm{b}}$  can be respectively defined as
\begin{align}
&h^-(\omega) = \left\{y \in \Phi_{\mathrm{b}}':  \mathcal{V}_{0, y}^2 \subseteq \mathcal{C}_0^{\mathrm{E}}, \quad y \in \Phi_{\mathrm{b}}'  \right\},\\
&H^-\left( Y \right) = Y + h^-\left( T_Y \right).
\end{align}
Using the notations above, we are able to characterize the relationship between $\lambda'$ and $\lambda$ by the mass transport formula \cite[Theorem 4.3.1]{BacBla:09}
\begin{align}
\lambda' \mathbb{E}^{0'}\left[ \mathrm{card}\left( H^+(0) \right) \right] = \lambda \mathbb{E}^{0}\left[ \mathrm{card}\left( H^-(0) \right) \right]
\end{align}
where $\mathrm{card}(\cdot)$ is the cardinality of point's set of neighbors. Since $2\mathrm{card}\left( H^+(0) \right) =   \mathrm{card}\left( H^-(0) \right)$, it is $\lambda' = 2 \lambda$. As such, if we denote $\mathcal{V}$ as a typical Voronoi cell of $\Phi_{\mathrm{b}}$ and $\mathcal{V}'$ as a typical Voronoi cell of $\Phi_{\mathrm{b}}'$, the following holds by using the mass transport theorem \cite{BacBla:09}
\begin{align}
\lambda \mathbb{E}^{0'}\left[ \left\vert V \circ H^-(0) \right\vert \right] = \lambda' \mathbb{E}^{0}\left[ \left\vert V' \circ H^+(0) \right\vert \right]
\end{align}
where $|\cdot|$ denotes the Lebesgue measure and $\circ$ denotes the composition operation. The above can be equivalently read as
\begin{align}\label{equ:AreaRelation}
\mathbb{E}\left[ \left\vert \mathcal{C}_i^{\mathrm{E}} \right\vert \right] =  2 \mathbb{E}\left[ \left\vert \mathcal{C}_i \right\vert \right].
\end{align}
As a result, the expectation of $K'$ can be calculated as
\begin{align}
\mathbb{E}\left[ K' \right]
&\stackrel{(a)}{=} \lambda_{\mathrm{u}} \mathbb{E}\left[\left| \mathcal{C}_{i}^{\mathrm{E}} \right| \right] - \lambda_{\mathrm{u}} \mathbb{E}\left[\left| \mathcal{C}_{i}  \right| \right]
= K
\end{align}
where (a) follows from the fact that mean number of UEs in area $\mathcal{A}$ is given by $\lambda_{\mathrm{u}}\vert \mathcal{A} \vert$.

\subsection{Proof of Corollary~\ref{cor:special_case}}\label{apx:special_case}
We start with the asymptotic result for the network coverage probability under CEU-ZF. When $\tau^2 = 0$, $\bar{\tau}^2 = 0$, and $\alpha=4$,
using Fubini's theorem \cite[Theorem 18.3]{Bil:08}, the coverage probability in \eqref{equ:CovProb_SCP} can be written as
\begin{align}\label{equ:cov_asym_ceuzf}
&\mathbb{P}\left( \gamma_{\mathrm{u},ik} \geq \theta \right)
\nonumber\\
&\approx \frac{1}{\Gamma(\mu)} \int_0^\infty\!\!\! \exp\!\left( - \left( \frac{\Omega}{\mu} t \right)^{\frac{1}{\eta}}\! \frac{ \pi (\lambda \pi)^2 \theta}{(1-\beta) N} \right) t^{\mu-1} e^{-t} dt.
\end{align}
As $\beta \ll 1$, the exponential term in \eqref{equ:cov_asym_ceuzf} can be approximated by its first order Taylor expansion, as follows
\begin{align}\label{equ:exp_ceuzf}
\exp\left( -  \frac{ \pi (\lambda \pi)^2 \theta \left( {\Omega} t /{\mu}  \right)^{\frac{1}{\eta}} }{(1-\beta) N} \right)
\approx 1 - \frac{ \pi (\lambda \pi)^2 \theta \left( {\Omega} t /{\mu}  \right)^{\frac{1}{\eta}}  }{(1-\beta) N}.
\end{align}
By substituting \eqref{equ:exp_ceuzf} into \eqref{equ:cov_asym_ceuzf}, the coverage probability under CEU-ZF reads as
\begin{align}\label{equ:cov_asym_ceuzf2}
\mathbb{P}\left( \gamma_{\mathrm{u},ik} \geq \theta \right) \approx 1 - \frac{\Gamma\left( \frac{1}{\eta} + \mu \right)}{\Gamma\left(  \mu \right)} \left( \frac{\Omega}{\mu} t \right)^{\frac{1}{\eta}} \frac{\left( \lambda \pi \right)^2 \theta }{(1-\beta)N},
\end{align}
and \eqref{equ:CovProb_CEUZF_Asym} follows from using \eqref{equ:Omega_Expression} and \eqref{equ:ERk} into \eqref{equ:cov_asym_ceuzf2}.

Similarly, using Fubini's theorem \cite{Bil:08},
we are able to approximate the coverage probability in \eqref{equ:OP_GenKm} as follows
\begin{align}\label{equ:cov_asym_ceazf}
\mathbb{P}\left( \gamma_{\mathrm{a},ik} \geq \theta \right) &\approx \frac{1}{\Gamma(\mu)} \int_0^\infty \!\! \left( 1 \!+\!  \frac{\left( {\Omega} t / {\mu} \right)^{\frac{1}{\eta}} (\lambda \pi)^2 \theta}{(1-2\beta)N} \right)
\nonumber\\
&\times\exp\left( - \frac{\left( {\Omega} t / {\mu} \right)^{\frac{1}{\eta}} (\lambda \pi)^2 \theta}{(1-2\beta)N}  \right) t^{\mu - 1} e^{-t} dt.
\end{align}
For $\beta \ll 1$, we have
\begin{align}\label{equ:exp_ceazf}
\exp\left( - \frac{\left( {\Omega} t / {\mu} \right)^{\frac{1}{\eta}} (\lambda \pi)^2 \theta}{(1-2\beta)N}  \right) \approx 1 - \frac{\left( {\Omega} t / {\mu} \right)^{\frac{1}{\eta}} (\lambda \pi)^2 \theta}{(1-2\beta)N}.
\end{align}
By substituting \eqref{equ:exp_ceazf} into \eqref{equ:cov_asym_ceazf}, the coverage probability under CEA-ZF reads as
\begin{align}\label{equ:cov_asym_ceazf2}
\mathbb{P}\left( \gamma_{\mathrm{a},ik} \geq \theta \right) \approx 1 - \frac{\Gamma\left( \frac{2}{\eta} + \mu \right)}{\Gamma\left(  \mu \right)} \left( \frac{\Omega}{\mu} t \right)^{\frac{2}{\eta}} \frac{\left( \lambda \pi \right)^4 \theta^2 }{(1-2\beta)^2 N^2},
\end{align}
and \eqref{equ:CovProb_CEAZF_Asym} follows from using \eqref{equ:Omega_Expression} and \eqref{equ:Rk2} into \eqref{equ:cov_asym_ceazf2}.
\end{appendix}
\balance
\bibliographystyle{IEEEtran}
\bibliography{bib/StringDefinitions,bib/IEEEabrv,bib/bscoop}

\begin{thebibliography}{10}
\providecommand{\url}[1]{#1}
\csname url@samestyle\endcsname
\providecommand{\newblock}{\relax}
\providecommand{\bibinfo}[2]{#2}
\providecommand{\BIBentrySTDinterwordspacing}{\spaceskip=0pt\relax}
\providecommand{\BIBentryALTinterwordstretchfactor}{4}
\providecommand{\BIBentryALTinterwordspacing}{\spaceskip=\fontdimen2\font plus
\BIBentryALTinterwordstretchfactor\fontdimen3\font minus
  \fontdimen4\font\relax}
\providecommand{\BIBforeignlanguage}[2]{{%
\expandafter\ifx\csname l@#1\endcsname\relax
\typeout{** WARNING: IEEEtran.bst: No hyphenation pattern has been}%
\typeout{** loaded for the language `#1'. Using the pattern for}%
\typeout{** the default language instead.}%
\else
\language=\csname l@#1\endcsname
\fi
#2}}
\providecommand{\BIBdecl}{\relax}
\BIBdecl

\bibitem{Ericsson:16}
{Ericsson}, ``{5G} radio access - {C}apabilities and technologies,''
  \emph{white paper}, Apr. 2016.

\bibitem{Nokia:15}
{Nokia Networks}, ``Ten key rules of {5G} deployment - {E}nabling
  1~{T}bit/s/km$^2$ in 2030,'' \emph{white paper}, Apr. 2015.

\bibitem{AndBuzCho:14}
J.~G. Andrews, S.~Buzzi, W.~Choi, S.~V. Hanly, A.~Lozano, A.~C. Soong, and
  J.~C. Zhang, ``What will {5G} be?'' \emph{{IEEE} J. Sel. Areas Commun.},
  vol.~32, no.~6, pp. 1065--1082, Jun. 2014.

\bibitem{Mar:10}
T.~L. Marzetta, ``Noncooperative cellular wireless with unlimited numbers of
  base station antennas,'' \emph{{IEEE} Trans. Wireless Commun.}, vol.~9,
  no.~11, pp. 3590--3600, Nov. 2010.

\bibitem{RusPerBuoLar:2013}
F.~Rusek, D.~Persson, B.~K. Lau, E.~G. Larsson, T.~L. Marzetta, O.~Edfors, and
  F.~Tufvesson, ``Scaling up {MIMO}: Opportunities and challenges with very
  large arrays,'' \emph{IEEE Signal Process. Mag.}, vol.~30, no.~1, pp. 40--60,
  Oct. 2013.

\bibitem{LuLiSwi:14}
L.~Lu, G.~Y. Li, A.~L. Swindlehurst, A.~Ashikhmin, and R.~Zhang, ``An overview
  of massive {MIMO}: Benefits and challenges,'' \emph{IEEE J. Sel. Topics
  Signal Process.}, vol.~8, no.~5, pp. 742--758, Oct. 2014.

\bibitem{NigMinHae:14}
G.~Nigam, P.~Minero, and M.~Haenggi, ``Coordinated multipoint joint
  transmission in heterogeneous networks,'' \emph{{IEEE} Trans. Commun.},
  vol.~62, no.~11, pp. 4134--4146, Nov. 2014.

\bibitem{XuYanLi:14}
Z.~Xu, C.~Yang, G.~Y. Li, Y.~Liu, and S.~Xu, ``Energy-efficient {CoMP}
  precoding in heterogeneous networks,'' \emph{{IEEE} Trans. Signal Process.},
  vol.~62, no.~4, pp. 1005--1017, Feb. 2014.

\bibitem{RalSinAnd:14}
R.~Tanbourgi, S.~Singh, J.~G. Andrews, and F.~K. Jondral, ``A tractable model
  for noncoherent joint-transmission base station cooperation,'' \emph{{IEEE}
  Trans. Wireless Commun.}, vol.~13, no.~9, pp. 4959--4973, Sep. 2014.

\bibitem{LozHeaAnd:13}
A.~Lozano, {R.~W.~Heath~Jr.}, and J.~G. Andrews, ``Fundamental limits of
  cooperation,'' \emph{{IEEE} Trans. Inf. Theory}, vol.~59, no.~9, pp.
  5213--5226, Sep. 2013.

\bibitem{ZakHan:12}
R.~Zakhour and S.~V. Hanly, ``Base station cooperation on the downlink: Large
  system analysis,'' \emph{{IEEE} Trans. Inf. Theory}, vol.~58, no.~4, pp.
  2079--2106, Apr. 2012.

\bibitem{BhaHea:11}
R.~Bhagavatula and {R.~W.~Heath~Jr.}, ``Adaptive limited feedback for sum-rate
  maximizing beamforming in cooperative multicell systems,'' \emph{{IEEE}
  Trans. Signal Process.}, vol.~59, no.~2, pp. 800--811, Jan. 2011.

\bibitem{HuaDurZho:15}
Y.~Huang, S.~Durrani, and X.~Zhou, ``Interference suppression using generalized
  inverse precoder for downlink heterogeneous networks,'' \emph{IEEE Wireless
  Commun. Lett.}, vol.~4, no.~3, pp. 325--328, Jun. 2015.

\bibitem{HoyHosBri:13}
J.~Hoydis, K.~Hosseini, S.~ten Brink, and M.~Debbah, ``Making smart use of
  excess antennas: Massive {MIMO}, small cells, and {TDD},'' \emph{Bell Labs
  Tech. J.}, vol.~18, no.~2, pp. 5--21, Sep. 2013.

\bibitem{BjoLarDeb:16}
E.~Bj{\"o}rnson, E.~G. Larsson, and M.~Debbah, ``Massive {MIMO} for maximal
  spectral efficiency: How many users and pilots should be allocated?''
  \emph{{IEEE} Trans. Wireless Commun.}, vol.~15, no.~2, pp. 1293--1308, Feb.
  2016.

\bibitem{ZhuWanQia:16}
X.~Zhu, Z.~Wang, C.~Qian, L.~Dai, J.~Chen, S.~Chen, and L.~Hanzo, ``Soft pilot
  reuse and multi-cell block diagonalization precoding for massive {MIMO}
  systems,'' \emph{{IEEE} Trans. Veh. Technol.}, vol.~PP, no.~99, pp. 1--1,
  2015.

\bibitem{hoy:13massive}
J.~Hoydis, S.~ten Brink, and M.~Debbah, ``Massive {MIMO} in the {UL}/{DL} of
  cellular networks: How many antennas do we need?'' \emph{{IEEE} J. Sel. Areas
  Commun.}, vol.~31, no.~2, pp. 160--171, Feb. 2013.

\bibitem{Fujitsu:11}
{Fujitsu Network Communications}, ``Enhancing {LTE} cell-edge performance via
  {PDCCH ICIC},'' \emph{white paper}, Mar. 2011.

\bibitem{BjoLar:15}
E.~Bj{\"o}rnson and E.~G. Larsson, ``Three practical aspects of massive {MIMO}:
  Intermittent user activity, pilot synchronism, and asymmetric deployment,''
  in \emph{Proc. IEEE Global Telecomm. Conf.}, San Diego, CA, Dec. 2015, pp.
  1--6.

\bibitem{TayDhiNov:12}
D.~B. Taylor, H.~S. Dhillon, T.~D. Novlan, and J.~G. Andrews, ``Pairwise
  interaction processes for modeling cellular network topology,'' in
  \emph{Proc. IEEE Global Telecomm. Conf.}, Anaheim, CA, Dec. 2012, pp.
  4524--4529.

\bibitem{BlaKarKee:13}
B.~Blaszczyszyn, M.~K. Karray, and H.~P. Keeler, ``Using {P}oisson processes to
  model lattice cellular networks,'' in \emph{Proc. IEEE Conf. on Computer
  Commun.}\hskip 1em plus 0.5em minus 0.4em\relax IEEE, Apr. 2013, pp.
  773--781.

\bibitem{YanGerQue:16}
H.~H. Yang, G.~Geraci, and T.~Q.~S. Quek, ``Energy-efficient design of {MIMO}
  heterogeneous networks with wireless backhaul,'' \emph{IEEE Trans. Wireless
  Commun.}, vol.~5, no.~7, pp. 4914--4927, Jul. 2016.

\bibitem{SinZhaAnd:15}
S.~Singh, X.~Zhang, and J.~G. Andrews, ``Joint rate and {SINR} coverage
  analysis for decoupled uplink-downlink biased cell associations in
  {HetNets},'' \emph{IEEE Trans. Wireless Commun.}, vol.~14, no.~10, pp.
  5360--5373, Oct. 2015.

\bibitem{Lee:82}
D.-T. Lee, ``On k-nearest neighbor {Voronoi} diagrams in the plane,''
  \emph{IEEE Trans. Comput.}, vol.~31, no.~6, pp. 478--487, Jun. 1982.

\bibitem{BacGio:15}
F.~Baccelli and A.~Giovanidis, ``A stochastic geometry framework for analyzing
  pairwise-cooperative cellular networks,'' \emph{{IEEE} Trans. Wireless
  Commun.}, vol.~14, no.~2, pp. 794--808, Feb. 2015.

\bibitem{LopGuvChu:12}
D.~Lopez-Perez, I.~Guvenc, and X.~Chu, ``Mobility management challenges in
  {3GPP} heterogeneous networks,'' \emph{IEEE Commun. Mag.}, vol.~50, no.~12,
  pp. 70--78, Dec. 2012.

\bibitem{CouDeb:11}
R.~Couillet and M.~Debbah, \emph{Random matrix methods for wireless
  communications}.\hskip 1em plus 0.5em minus 0.4em\relax Cambridge University
  Press, 2011.

\bibitem{WagCouDeb:12}
S.~Wagner, R.~Couillet, M.~Debbah, and D.~T. Slock, ``Large system analysis of
  linear precoding in correlated {MISO} broadcast channels under limited
  feedback,'' \emph{{IEEE} Trans. Inf. Theory}, vol.~58, no.~7, pp. 4509--4537,
  Mar. 2012.

\bibitem{GerAlnYua:13}
G.~Geraci, A.~Y. Al-nahari, J.~Yuan, and I.~B. Collings, ``Linear precoding for
  broadcast channels with confidential messages under transmit-side channel
  correlation,'' \emph{IEEE Commun. Lett.}, vol.~17, no.~6, pp. 1164--1167, May
  2013.

\bibitem{MulCotVeh:2014}
R.~R. M{\"u}ller, L.~Cottatellucci, and M.~Vehkapera, ``Blind pilot
  decontamination,'' \emph{IEEE J. Sel. Topics Signal Process.}, vol.~8, no.~5,
  pp. 773--786, 2014.

\bibitem{FerGerQue:16}
G.~C. Ferrante, G.~Geraci, T.~Q.~S. Quek, and M.~Z. Win, ``Group-blind
  detection for uplink of massive {MIMO} systems,'' \emph{{IEEE} Trans. Signal
  Process.}, 2016, to appear.

\bibitem{MarHoc:06}
T.~L. Marzetta and B.~M. Hochwald, ``Fast transfer of channel state information
  in wireless systems,'' \emph{{IEEE} Trans. Signal Process.}, vol.~54, no.~4,
  pp. 1268--1278, Apr. 2006.

\bibitem{BjoLarMar:15}
E.~Bj{\"o}rnson, E.~G. Larsson, and T.~L. Marzetta, ``Massive {MIMO}: Ten myths
  and one critical question,'' \emph{IEEE Commun. Mag.}, vol.~54, no.~2, pp.
  114--123, Feb. 2016.

\bibitem{MulCouBjo:15}
A.~M{\"u}ller, R.~Couillet, E.~Bj{\"o}rnson, S.~Wagner, and M.~Debbah,
  ``Interference-{Aware} {RZF} precoding for multicell downlink systems,''
  \emph{{IEEE} Trans. Signal Process.}, vol.~63, no.~15, pp. 3959--3973, Apr.
  2015.

\bibitem{GerCouYua:13}
G.~Geraci, R.~Couillet, J.~Yuan, M.~Debbah, and I.~B. Collings, ``Large system
  analysis of linear precoding in {MISO} broadcast channels with confidential
  messages,'' \emph{{IEEE} J. Sel. Areas Commun.}, vol.~31, no.~9, pp.
  1660--1671, Aug. 2013.

\bibitem{NosBehAnd:11}
B.~Nosrat-Makouei, J.~G. Andrews, and R.~W. Heath, ``{MIMO} interference
  alignment over correlated channels with imperfect {CSI},'' \emph{{IEEE}
  Trans. Signal Process.}, vol.~59, no.~6, pp. 2783--2794, Jun. 2011.

\bibitem{WanMur:06}
C.~Wang and R.~D. Murch, ``Adaptive downlink multi-user mimo wireless systems
  for correlated channels with imperfect {CSI},'' \emph{{IEEE} Trans. Wireless
  Commun.}, vol.~5, no.~9, pp. 2435--2446, Sep. 2006.

\bibitem{GerEgaYua:12}
G.~Geraci, M.~Egan, J.~Yuan, A.~Razi, and I.~B. Collings, ``Secrecy sum-rates
  for multi-user {MIMO} regularized channel inversion precoding,'' \emph{{IEEE}
  Trans. Commun.}, vol.~60, no.~11, pp. 3472--3482, Nov. 2012.

\bibitem{PeeHocSwi:05}
C.~B. Peel, B.~M. Hochwald, and A.~L. Swindlehurst, ``A vector-perturbation
  technique for near-capacity multiantenna multiuser communication-part {I}:
  channel inversion and regularization,'' \emph{{IEEE} Trans. Commun.},
  vol.~53, no.~1, pp. 195--202, Feb. 2005.

\bibitem{KamMulBjo:2014}
A.~Kammoun, A.~M{\"u}ller, E.~Bj{\"o}rnson, and M.~Debbah, ``Linear precoding
  based on polynomial expansion: Large-scale multi-cell {MIMO} systems,''
  \emph{IEEE J. Sel. Topics Signal Process.}, vol.~8, no.~5, pp. 861--875, Jan.
  2014.

\bibitem{BacBla:09}
F.~Baccelli and B.~Blaszczyszyn, \emph{Stochastic Geometry and Wireless
  Networks. Volumn I: Theory}.\hskip 1em plus 0.5em minus 0.4em\relax Now
  Publishers, 2009.

\bibitem{GioDhiAnd:14}
G.~Geraci, H.~S. Dhillon, J.~G. Andrews, J.~Yuan, and I.~B. Collings,
  ``Physical layer security in downlink multi-antenna cellular networks,''
  \emph{{IEEE} Trans. Commun.}, vol.~62, no.~6, pp. 2006--2021, Jun. 2014.

\bibitem{HeaKouBai:13}
{R.~W.~Heath~Jr.}, M.~Kountouris, and T.~Bai, ``Modeling heterogeneous network
  interference using {Poisson} point processes,'' \emph{{IEEE} Trans. Signal
  Process.}, vol.~61, no.~16, pp. 4114--4126, Aug. 2013.

\bibitem{hae:12}
M.~Haenggi, \emph{Stochastic geometry for wireless networks}.\hskip 1em plus
  0.5em minus 0.4em\relax Cambridge University Press, 2012.

\bibitem{Bil:08}
P.~Billingsley, \emph{Probability and measure}.\hskip 1em plus 0.5em minus
  0.4em\relax John Wiley \& Sons, 2008.

\bibitem{FilYac:06}
J.~C.~S. Santos~Filho and M.~D. Yacoub, ``Simple precise approximations to
  {W}eibull sums,'' \emph{{IEEE} Commun. Lett.}, vol.~10, no.~8, pp. 614--616,
  Aug. 2006.

\bibitem{DhiKouAnd:13}
H.~S. Dhillon, M.~Kountouris, and J.~G. Andrews, ``Downlink {MIMO} hetnets:
  Modeling, ordering results and performance analysis,'' \emph{{IEEE} Trans.
  Wireless Commun.}, vol.~12, no.~10, pp. 5208--5222, Oct. 2013.

\bibitem{LiZhaLet:14}
C.~Li, J.~Zhang, and K.~Letaief, ``Throughput and energy efficiency analysis of
  small cell networks with multi-antenna base stations,'' \emph{{IEEE} Trans.
  Wireless Commun.}, vol.~13, no.~5, pp. 2505 -- 2517, May 2014.

\bibitem{LiBjoLar:15}
X.~Li, E.~Bjornson, E.~G. Larsson, S.~Zhou, and J.~Wang, ``A multi-cell {MMSE}
  detector for massive {MIMO} systems and new large system analysis,'' in
  \emph{Proc. IEEE Global Telecomm. Conf.}\hskip 1em plus 0.5em minus
  0.4em\relax IEEE, Feb. 2015, pp. 1--6.

\bibitem{JosAshMar:11}
J.~Jose, A.~Ashikhmin, T.~L. Marzetta, and S.~Vishwanath, ``Pilot contamination
  and precoding in multi-cell {TDD} systems,'' \emph{{IEEE} Trans. Wireless
  Commun.}, vol.~10, no.~8, pp. 2640--2651, Jun. 2011.

\bibitem{AtzArnDeb:15}
I.~Atzeni, J.~Arnau, and M.~Debbah, ``Fractional pilot reuse in massive {MIMO}
  systems,'' in \emph{Proc. IEEE Int. Conf. Commun.}, London, UK, Jun. 2015,
  pp. 1030--1035.

\bibitem{HanTse:01}
S.~V. Hanly and D.~N.~C. Tse, ``Resource pooling and effective bandwidths in
  {CDMA} networks with multiuser receivers and spatial diversity,''
  \emph{{IEEE} Trans. Inf. Theory}, vol.~47, no.~4, pp. 1328--1351, May 2001.

\bibitem{DalVer:07}
D.~J. Daley and D.~Vere-Jones, \emph{An introduction to the theory of point
  processes: volume {II}: general theory and structure}.\hskip 1em plus 0.5em
  minus 0.4em\relax Springer Science \& Business Media, 2007.

\end{thebibliography}

\end{document}